\documentclass{gGAF2e} 

\usepackage[T1]{fontenc}
\usepackage{ae,aecompl}
\usepackage[varg]{txfonts}

\usepackage[utf8]{inputenc}

\usepackage{graphicx}
\usepackage[usenames,dvipsnames]{color}
\usepackage{xspace}
\usepackage{bm}
\graphicspath{./fig/} %path(s) to graphics files

\usepackage{ulem,cancel,xcolor}

\newcommand{\cm}{\,{\rm cm}}

\newcommand{\erg}{\,{\rm erg}}
\newcommand{\g}{\,{\rm g}}

\newcommand{\kms}{\,{\rm km\,s^{-1}}}

\newcommand{\K}{\,{\rm K}}
\newcommand{\kpc}{\,{\rm kpc}}
\newcommand{\p}{\,{\rm pc}}

\newcommand{\Myr}{\,{\rm Myr}}
\newcommand{\Gyr}{\,{\rm Gyr}}

\newcommand{\mkG}{\,\mu{\rm G}}

\newcommand{\Msol}{\,{\rm M_{\odot}}}

\newcommand{\yr}{\,{\rm yr}}
\newcommand{\ion}[2]{${\rm #1}{\rm #2}$}
\newcommand{\rsub}[2]{#1_{\mathrm{#2}}}

\newcommand{\vect}[1]{\boldsymbol{#1}}%vector
\newcommand{\dd}{\mathrm{d}}%differential
\newcommand{\ii}{\mathrm{i}}%imaginary unity
\newcommand{\mean}[1]{\langle{#1}\rangle}

\usepackage[pdftex,breaklinks=true,pdfborder={0 0 0.05},colorlinks]{hyperref}

\usepackage{natbib,twoopt}

\begin{document}

\jvol{00} \jnum{00} \jyear{2012} %\jmonth{February}

\markboth{J.F.HOLLINS, G.R.SARSON, C.C.EVIRGEN, A.SHUKUROV, A.FLETCHER AND F.A.GENT}{GEOPHYSICAL \& ASTROPHYSICAL FLUID DYNAMICS}

\title{
Mean fields and fluctuations in compressible magnetohydrodynamic flows}

\author{
James~F.~Hollins${\dag}$,
Graeme~R.~Sarson${\dag}$$^{\ast}$\thanks{$^\ast$Corresponding author. Email: graeme.sarson@newcastle.ac.uk \vspace{6pt}},
Cetin~Can~Evirgen${\dag}$,
Anvar~Shukurov${\dag}$,
Andrew~Fletcher${\dag}$ and 
Fred~A.~Gent${\ddag}$\\
\vspace{6pt}  
${\dag}$School of Mathematics, Statistics and Physics, Newcastle University, Newcastle upon Tyne, NE1 7RU, UK\\
${\ddag}$Astroinformatics, Department of Computer Science, Aalto University, PO Box 15400, FI-00076, Espoo, Finland\\
\vspace{6pt}\received{submitted October 2021} }

\maketitle

%-------------------------------------------------------------------------------

\begin{abstract} 
We apply Gaussian smoothing to obtain mean magnetic field, density, velocity, and magnetic and kinetic energy densities 
from our numerical model of the interstellar medium,
based on three-dimensional magnetohydrodynamic equations
in a shearing box $1\times1\times2\kpc$ in size. 
The interstellar medium is highly compressible, as the turbulence is transonic or supersonic;
it is thus an excellent context in which to explore the use of smoothing 
to represent physical variables in a compressible medium 
in terms of their mean and fluctuating parts.
Unlike alternative averaging procedures, such as horizontal averaging, 
Gaussian smoothing retains the three-dimensional structure of the mean fields. 
Although Gaussian smoothing does not obey the Reynolds rules of averaging, 
physically meaningful and mathematically sound central statistical moments 
are defined as suggested by \citet{Germano:1992}.
We discuss methods to identify an optimal smoothing scale $\ell$ and the effects of this choice on the results. 
From spectral analysis of the magnetic, density and velocity fields, 
we find a suitable smoothing length for all three fields, of $\ell \approx 75\p$.
Such a smoothing scale is likely to be sensitive to the choice of simulation 
parameters; this may be considered in future work, 
but here we just explore the methodology.
We discuss the properties of third-order statistical moments in fluctuations of kinetic energy density in compressible flows, and suggest their physical interpretation. 
The mean magnetic field, amplified by a mean-field dynamo, significantly alters the distribution of kinetic energy in space and between scales, 
reducing the magnitude of kinetic energy at intermediate scales.
This intermediate-scale kinetic energy is a useful diagnostic of the importance of SN-driven outflows.
\end{abstract}


%-------------------------------------------------------------------------------

\section{Introduction}\label{sect:intro}

The injection of thermal and kinetic energy by stellar winds and supernova (SN)
explosions drives transonic turbulence in the interstellar medium (ISM) and 
produces an inhomogeneous, multiphase system
\citep{Elmegreen:2004,Scalo:2004,MacLow:2004}. 
The outer scale of turbulent motions in the ISM consistently suggested by 
observations, theory and simulations is of order $10\text{--}100\p$, and the
turbulent scales extend to a fraction of a parsec \citep{Armstrong:1995}.

Understanding the properties and nature of a turbulent flow requires the
separation of mean and fluctuating quantities. Such a separation is well
understood for statistically homogeneous random flows where a number of 
averaging procedures are available. Volume or area averaging are most important 
in astronomy, while numerical simulations provide a further opportunity to 
average over time. Under favourable conditions (defined by ergodic theorems 
and hypotheses), the resulting averages are equivalent to the statistical ensemble 
averages employed in theory 
\citep[e.g.,][]{Monin:1975,Panchev:1971,Tennekes:1972}. The 
ensemble averages are rarely accessible in applications, 
as their calculation requires the availability of a large number 
of statistically independent realizations of the random processes.

Space and time averaging procedures are consistent with ensemble averaging 
$\mean{f+g}=\mean{f}+\mean{g}$, $\mean{\mean{f}g}=\mean{f}\mean{g}$  
and $\mean{\mean{f}}=\mean{f}$,
where $f$ and $g$ are random functions and angular brackets denote averaging 
\citep[e.g., Sect.~3.1 in][]{Monin:1975}. Volume and time averaging only satisfy 
the Reynolds rules
in an approximate manner when the scales of 
variations of the mean quantities and the fluctuations differ significantly (the 
requirement of scale separation between the averaged quantities and the
fluctuations) and the averaging scale is large in comparison with the scale of
the fluctuations and small in comparison with that of the mean quantities. In
practice, the mean quantities need to be homogeneous or time-independent for
the ensemble and volume (or time) averages to be consistent with each other.

The outer scale of the interstellar turbulence is comparable to the scale 
height of the gas density distribution in spiral galaxies (about $0.1\kpc$ and 
$0.5\kpc$ for the cold and warm diffuse \ion{H}{I}, respectively). Therefore, 
the interstellar turbulent flow cannot be considered statistically homogeneous
apart from along the horizontal directions. However, numerical simulations of 
the supernova-driven, multi-phase ISM have relatively small horizontal domains
of order $1\kpc\times1\kpc$ or less   
\citep[e.g.,][]{Korpi:1999b, Joung:2006, deAvillez:2007, deAvillez:2012a, 
deAvillez:2012b, Gressel:2008, Federrath:2010, Hill:2012, Gent:2013b, Gent:2013a, 
Gressel:2013, Bendre:2015, Walch:2015, Girichidis:2016a, Girichidis:2016b, 
Girichidis:2018a}. 
Meanwhile, the ISM has a wide range of density and velocity structures (e.g., 
those related to gas clouds, galactic outflows and spiral patterns) that cover 
continuously the range of scales from $1 \p$ to $10\kpc$. Therefore, scale 
separation between the random and large-scale ISM flows is questionable at best. 
This poses difficulties for the interpretation of numerical simulations. Similar 
difficulties arise in the interpretation of observations, but numerical simulations 
have exposed the problems especially clearly.

%The
MHD simulations of galaxies
employ domains of scale 
$10\kpc$ and larger,
and contain kiloparsec-scale structures that would be associated with 
mean-field dynamics \citep[e.g.,][]{Slyz:2003, Dobbs:2008, Hanasz:2009, 
Kulesza-Zydzik:2009, Dobbs:2010, Pakmor:2016, Pakmor:2017, Rieder:2016, Rieder:2017}. 
Such simulations can also support a small-scale magnetic field driven by dynamo action
\citep[][]{Pakmor:2016, Pakmor:2017, Rieder:2016, Rieder:2017}. However, since 
computational limits restrict the resolution
of such simulations to order $10 \p$, 
these simulations are unable to fully model physics such as the injection of SN energy 
into the ISM and must rely on a simplified prescription. Thus, whilst scale separation 
may be obtained and lead to useful results for the magnetic field, such results for the 
velocity field may not be reliable.

Simulations on the scale of interstellar clouds have provided powerful insight
into the parsec and sub-parsec physics in the ISM \citep[e.g.,][]{Klessen:2000, 
Heitsch:2001, Brandenburg:2007}. Such simulations drive turbulence at specific 
wavenumbers to simulate ISM turbulence. Whilst scale separation may be possible,
it would be a separation between the parameterised injected flow and any resultant
smaller-scale flows. Thus, the results from these scale separations may not be 
fully informative.

The division of the Navier--Stokes and magnetohydrodynamic (MHD) equations into 
evolution equations for the mean flow and the fluctuations has been explored for 
both ensemble averaging and filtering of the fluctuations (also known as 
coarse-graining); i.e., volume averaging via convolution with a compact kernel. 
The Reynolds rules of averaging are not satisfied for this procedure but this is 
not an obstacle to developing a mathematically sound formalism that leads to 
evolution equations for averaged quantities and their moments \citep{Germano:1992}. 
The most widely known application of this technique is to subgrid models for large 
eddy simulations of turbulent flows \citep{Meneveau:2012}. \citet{Eyink:2018} and 
\citet{Aluie:2017} provide details and a review of this approach to  hydrodynamic 
and MHD turbulence \citep[see also][]{Eyink:1995,Eyink:2015}.

An important advantage of the filtering approach is that, together with 
ensemble averaging, it does not require scale separation between the mean 
fields and their fluctuations \citep[e.g.,][]{Aluie:2017}.
The separation of the mean and fluctuating quantities in a random flow is of
crucial significance in the theory of mean-field turbulent dynamos.
Mean-field dynamo theory is based on ensemble averaging, but
numerical simulations rely on various volume and time averaging procedures.
(We note that the algebraic form of the mean-field equations is the same for a wide range of averaging methods: \mbox{\citealp{Germano:1992,Aluie:2017}}.)
For example, the separation of the magnetic field into mean and fluctuating
components often involves averaging over the whole computational volume or, in
systems stratified along the $z$-direction due to gravity, averaging in the
$(x,y)$-planes \citep[horizontal averaging; see][]{Brandenburg:2005}. 
The resulting mean magnetic field is either perfectly uniform 
or only dependent on $z$. 
These constraints on the form of the mean magnetic field 
are often
artificial and unphysical. An inhomogeneous system, such as the ISM, is
expected to produce a spatially complex mean field, which is ignored by these
simple volume or horizontal averaging techniques. A further
complication with horizontal averaging, when periodic boundary conditions are
used in $x$ and $y$, is that $\langle B_z\rangle$ must vanish to guarantee
the solenoidality of the mean magnetic field
\citep[e.g.,][]{Gent:2013a}. 
The main advantage of horizontal averaging is to obey the Reynolds rules,
achieved often at the expense of physical validity.  
Another, Reynolds rules compliant,
option is to use azimuthal averaging in global simulations of dynamo action in
a rotating spherical object to obtain an axially symmetric mean magnetic field
\citep[see][for a
review]{Simard:2016}. This approach is easier to justify but still it excludes
physically admissible azimuthal variations of the mean field.
Furthermore, the kinematic mean-field dynamo 
action, with homogeneous transport coefficients $\alpha$ and $\beta$
(representing the $\alpha$-effect and turbulent diffusion), 
in infinite space produces an inhomogeneous mean magnetic field that varies at all 
wavenumbers below $\alpha/\beta$, with the dominant mode having the 
wavenumber $\alpha/(2\beta)$ \citep[e.g.,][]{Sokoloff:1983}. 
The spatial structure of any mean field is controlled 
by the physical properties of the system, 
rather than by the size of the computational domain.

We discuss an alternative approach to averaging based on Gaussian smoothing 
as suggested by \citet{Germano:1992}, and employ it to obtain 
the mean fields in simulations of the multi-phase, supernova-driven ISM. Averaging 
with a Gaussian (or another) kernel is inherent in astronomical observations, 
where such smoothing is applied either during data reduction or stems from the 
finite width of a telescope beam. This approach has been applied by 
\citet{Gent:2013a} to the simulated magnetic field; here we extend it to the 
velocity and density fields and, importantly, energy densities, which represent 
higher-order statistical moments. In particular, kinetic energy density in a 
compressible flow represents a third-order statistical moment and requires 
special attention. 

A detailed exploration of the sensitivity of our results to 
the parameter space is beyond the scope of this paper. Rather, we 
seek to demonstrate that the extension of this approach to also
concern the density and velocity, and thus also kinetic energy,
does indeed lead to mathematically and physically meaningful
results.

A summary of our numerical model of the ISM is presented in 
Section~\ref{sect:model_summary}, and Section~\ref{sect:mean_fields} introduces  
averaging based on Gaussian smoothing. Various approaches to the 
selection of the smoothing length are discussed in 
Section~\ref{sect:power_spectra}. Section~\ref{sect:energy_densities} analyses 
the behaviour of magnetic and kinetic energy densities. 
Section~\ref{sect:dynamo} details the effects of the amplified mean field on
the magnetic and kinetic energies.
Section~\ref{sect:horizontal} compares Gaussian smoothing with horizontal averaging,
to show the advantages of the former in the current context.
Finally, section~\ref{sect:discussion} summarises and concludes our discussion.

%-------------------------------------------------------------------------------

%\section{Simulations of the multiphase ISM}\label{sect:model_summary}
\section{A numerical model of the multiphase ISM}\label{sect:model_summary}

We use our earlier numerical model of the ISM, described in detail by
\citet{Gent:2013a,Gent:2013b}. The model involves solving,
with the Pencil Code \citep{JOSS21},
the full,
compressible, non-ideal MHD equations with parameters generally typical of the
Solar neighbourhood in a three-dimensional Cartesian, shearing box with radial
($x$) and azimuthal ($y$) extent $L_x=L_y=1.024\kpc$ 
and vertical ($z$) extent $L_z=1.086\kpc$ on either side of the midplane at $z=0$.
Our numerical resolution is $\Delta=\Delta x
= \Delta y = \Delta z = 4 \p$, using $256$ grid points in $x$ and $y$ and
$544$ in $z$. \citet{Gent:2013a} and, in greater detail, \citet{GMKSH20}
demonstrate that this resolution is
sufficient to reproduce the known solutions for expanding SN remnants in the
Sedov--Taylor and momentum-conserving phases.  Details of the numerical
implementation and its comparison with some other similar simulations can be
found in Appendix~\ref{sect:parameters}.

The mass conservation, Navier--Stokes, heat and induction equations are solved
for mass density $\rho$, velocity $\vect{u}$, specific entropy $s$, and
magnetic  vector potential $\vect{A}$ (such that
$\vect{B}=\nabla\times\vect{A}$).  The Navier--Stokes equation includes a
fixed vertical gravity force with  contributions from the stellar 
disc
and
dark halo. The initial state is an approximate hydrostatic equilibrium. The
Galactic differential rotation is modelled by a background shear flow
$\vect{U} = (0,-q \Omega x, 0)$, where $q$ is the shear parameter and $\Omega$
is the Galactic angular velocity.  Here we use $q=1$, as in a flat rotation
curve (i.e., with the rotation speed independent of the cylindrical radius),
and $\Omega=50\kms\kpc^{-1}$, twice that of the Solar neighbourhood in
order to enhance the mean-field dynamo action and thus reduce the
computational time.
The velocity $\vect{u}$ is the perturbation velocity in the rotating frame,
that remains after the subtraction of the background shear flow from the total
velocity. However, it still contains a large-scale vertical component due to
an outflow driven by the SN activity.

Both Type II and Type I supernova explosions (SNe) are included in the
simulations.  These differ only in their vertical distribution and frequency.
The frequencies used correspond to those in the Solar neighbourhood. We
introduce Type II SNe at a mean rate per surface area of $\nu_{\rm II} = 25
\kpc^{-2} \Myr^{-1}$.  Type I SNe have a mean rate per surface area of
$\nu_{\rm I} = 4 \kpc^{-2} \Myr^{-1}$.  The SN sites
have uniform random distribution in the horizontal plane. 
Their vertical positions have Gaussian distributions
with scale heights $h_{\rm II}=0.09\kpc$ and $h_{\rm I} = 0.325 \kpc$ for
SN\,II and SN\,I, respectively. 

No spatial clustering of the SNe is included since the size of superbubbles
produced by SNe clustering are comparable to the horizontal size of the
computational domain.  Simulations in a domain of significantly larger size
are required to capture the effects of the SN clustering.
\citet{deAvillez:2007} include SN clustering in their simulations and obtain
the correlation scale of the random flows of $75\p$, comparable to those
obtained from the correlation analysis of this model
\citep[see][]{Hollins:2017}.  Each SN is initialised as an injection of
$0.5\times10^{51}\erg$ of thermal energy and a variable amount of kinetic
energy that depends on the local gas density fluctuations and the turbulent ambient velocities,
and has the mean net value $0.4\times10^{51}\erg$.

We include optically thin radiative cooling with a parameterised cooling
function. For $T<10^5\K$, we adopt a power-law fit to the `standard
equilibrium' pressure--density curve of \citet[][]{Wolfire:1995}, as given in
\citet[][]{Sanchez-Salcedo:2002}. For $T>10^5\K$, we use the cooling function
of \citet[][]{Sarazin:1987}. Photoelectric heating is also included as in
\citet[][]{Wolfire:1995}; it decreases with $|z|$ on a length scale comparable
to the gas scale height near the Sun. The system exhibits distinct hot, warm
and cold gas phases identifiable as peaks in the joint probability
distribution of the gas density and temperature.

Shock-capturing kinetic, thermal and magnetic diffusivities (in addition to  
background diffusivities) are included to resolve shock discontinuities and
maintain numerical stability in the Navier--Stokes, heat and induction 
equations. Periodic boundary conditions are used in $y$, and sheared-periodic 
boundary conditions in $x$. Open boundary conditions, permitting outflow and 
inflow, are used at the vertical boundaries at $z=\pm L_z$.
\citet{Gent:2013a,Gent:2013b} provide further details on the boundary 
conditions used and on the other implementations briefly described above.

Starting with a weak azimuthal magnetic field at the midplane,
the system is susceptible to the dynamo instability. Dynamo action can be 
identified with exponential growth of magnetic field saturating after about 
$1.4\Gyr$ at a level of $2.5\mkG$, comparable to observational estimates for 
the solar neighbourhood \citep{Gent:2013a}. The magnetic field has energy at a 
scale comparable to the size of the computational domain, suggesting a 
mean-field dynamo action \citep{Gent:2013b}.

We analyse snapshots in the range $0.8 \leq t \leq 1.725\Gyr$.
Three distinct temporal stages can be identified 
in the dynamo action and the magnetic field.
With magnetic energy low, compared to the thermal and kinetic energies,
$0.8 \leq t < 1.1\Gyr$ hosts the kinematic phase of the mean-field dynamo.
The dynamo adjusts itself to a non-linear stage at $1.1 \leq t < 1.45\Gyr$ as the
magnetic energy reaches approximate equipartition with kinetic energy of the
random flow.  Finally, at $1.45 \leq t \leq 1.725\Gyr$, the mean-field dynamo
saturates and the magnetic energy slightly exceeds the kinetic energy
\citep[see][]{Gent:2013b}.  Since the evolution of the magnetic field is
expected to significantly affect the structure of the gas density and
velocity, each stage is considered separately. The results are illustrated in
the figures shown below using the snapshot at $t=1.6\Gyr$.

%-------------------------------------------------------------------------------

\section{Mean fields and fluctuations in a compressible random flow}
\label{sect:mean_fields}\label{sect:filtering}

Averaging procedures can be used to represent a physical variable $f$  
as a superposition of its
mean $\langle f \rangle$ and fluctuations $f'$:
$f=\langle f \rangle + f'$.
Ensemble averaging is used in most theoretical contexts. 
Ensemble-averaged quantities do not need to be independent of any spatial or temporal variable. 
However, volume and time averaging are often the only options available in simulations and 
observations, and those averages clearly are independent of spatial and time variables, respectively.
Consider for example, the average over a volume $V$,
\begin{align}\label{eq:reynolds}
\langle f \rangle_{V} = \frac{1}{V}\int_{V} f(\vect{x}') \, \dd^{3}\vect{x'}\,.
\end{align}
It satisfies the Reynolds rules of averaging, including
\begin{align}
\left\langle f \langle g \rangle_{V}\right\rangle_{V} 
&= \langle f \rangle_{V} \langle g \rangle_{V}\,, 
\qquad
\left\langle \langle f \rangle_{V} \right\rangle_{V} = \langle f \rangle_{V}\,,
\label{eq:Reynolds_rules}\\
\intertext{leading to}
\langle f' \rangle_{V} &= 0\,, \qquad
\left\langle \langle f \right\rangle_{V} \, g' \rangle_{V} = 0\,,
\label{eq:conditions_1}
\end{align}
for the random variables $f$ and $g$.
This allows evolutionary equations for the central moments 
$\langle f' g' \rangle_{V}$, $\langle f' g' h' \rangle_{V}$
(with another random variable $h$), etc.,
to be derived by averaging the governing equations using relations, such as
for the velocity field $\vect{u}$
\citep[e.g.,][]{Monin:1975}, 
\begin{align}
\langle u'_{i} u'_{j} \rangle_{V}& = \langle u_{i} u_{j} \rangle_{V} - 
	\langle u_{i} \rangle_{V} \langle u_{j} \rangle_{V}\,, \notag  \\
\langle u'_{i}  u'_{j} u'_{k} \rangle_{V} &= 
	\langle u_{i} u_{j} u_{k} \rangle_{V} 
		-\langle u_i\rangle_V \langle u'_j u'_k\rangle_V 
				-\langle u_j \rangle_V  \langle u'_k u'_i\rangle_V	%\notag \\
%&
- \langle u_k \rangle_V  \langle u'_i u'_j \rangle_V 
		%- \langle u_i \rangle_V  \langle u_j \rangle_V \langle u_k\rangle_V\,,
		- \langle u_i \rangle_V  \langle u_j \rangle_V \langle u_k\rangle_V\,.
\label{eq:central_moments_equations}
\end{align}
%in the case of the velocity field $\vect{u}$.

In numerical simulations, $V$ is often the whole computational domain, or some
significant part of it, or a (thin) slice parallel to one of the coordinate
planes, as in averages over a horizontal plane ($x,y)$, or azimuthal averaging.
Such averages are constrained to be partially
or fully independent of position, in all three directions in the case of
volume averages, in two dimensions for horizontal averages and in the azimuth
for axial averages. As we discuss in Section~\ref{sect:intro}, these
constraints may be, and often are, unreasonably restrictive. Moreover,
any observational data obtained with a finite resolution represent a
convolution of the quantity observed with the telescope beam, and are free to
vary with position. It is therefore desirable to apply to numerical results an
averaging procedure, compatible with the observational procedures, in a manner
that does not impose unjustifiable restrictions on the averaged quantities.
This is the goal of this paper.

%-------------------------------------------------------------------------------

A local mean part of a random field $f(\vect{x})$, denoted $\langle 
f\rangle_\ell$, is obtained by spatial smoothing (filtering) of its 
fluctuations at scales $l<\ell$, with a certain \textit{smoothing length} 
$\ell$,  using a smoothing kernel $\mathcal{G}_\ell$:
\begin{equation}\label{eq:filter}
\langle f(\vect{x}) \rangle_\ell 
= \int_V f(\vect{x}')\mathcal{G}_\ell(\vect{x}-\vect{x}') \,\dd^{3}\vect{x}'\,, 
\end{equation}
where integration extends over the whole volume where $f(\vect{x})$ is defined. 
The filtering kernel is normalized, and assumed to be symmetric,
\begin{equation}\label{Gsym} 
\int_V \mathcal{G}_\ell(\vect{x}-\vect{x}')\,\dd^3\vect{x}' = 1\,, \qquad
\int_{V} \vect{x} G_\ell(\vect{x}) \,\dd^3\vect{x} = 0\,.
\end{equation}
To ensure that fluctuations in kinetic energy density are positive definite,
the kernel must be positive for all $\vect{x}$ \citep[][and references 
therein]{Aluie:2017}.
The fluctuation field is obtained as
\begin{equation}\label{fluct}
f'(\vect{x}) = f(\vect{x}) -\langle f(\vect{x}) \rangle_{\ell}\,,
\end{equation}
(with the link between the prime and the scale $\ell$ being understood).
This procedure retains the spatial structure of both the mean field and the 
fluctuations. We discuss below physically motivated choices for the smoothing 
length $\ell$.

Thus defined, the averaging procedure does not satisfy the Reynolds rules  
outlined in equations~\eqref{eq:Reynolds_rules} and \eqref{eq:conditions_1}. 
In particular, the mean of the fluctuations does not vanish, repeated averaging 
affects the mean field $\langle f(\vect{x}\rangle_\ell$, and the mean and 
fluctuating fields are not uncorrelated:
\begin{equation}\label{eq:conditions_filtering}
\langle f' \rangle_{\ell} \neq 0\,, \qquad
\langle \langle f \rangle_\ell \rangle_\ell \neq \langle f \rangle_\ell\,, 
\qquad 
\langle \langle f \rangle_\ell  f' \rangle_\ell \neq 0\,.
\end{equation}
As a consequence, the standard relations between statistical moments of total 
fields and their fluctuations, shown in 
equation~\eqref{eq:central_moments_equations}, are no longer valid.

To address these complications, \citet{Germano:1992} introduced generalised 
statistical moments $\mu(f, g)\,\ \mu(f, g, h)\,\ \ldots\,,$ of random fields 
$f(\vect{x})$,  $g(\vect{x})$ and $h(\vect{x})$ to ensure that the mathematical 
soundness and simplicity of the averaged governing equations is regained for 
both the mean fields and their statistical moments. In fact, relations between 
the statistical moments are quite similar to the standard ones of 
equation~\eqref{eq:central_moments_equations}. For example, the generalised 
statistical moments of the velocity field $\vect{u}(\vect{x})$ are defined as
\begin{align}\nonumber
\mu(u_i,u_j)&= \langle u_iu_j\rangle_\ell 
	-\langle u_i \rangle_\ell \langle u_j \rangle_\ell\,,\\\label{eq:generalised_moments_equations}
\mu(u_i, u_j, u_k)&= \langle u_i u_j u_k \rangle_\ell 
	- \langle u_i \rangle_\ell\, \mu (u_j, u_k) 
		- \langle u_j \rangle_\ell\, \mu (u_k, u_i) %\notag \\
%	&\mbox{}\quad
- \langle u_k \rangle_\ell\, \mu (u_i, u_j) 
		- \langle u_i \rangle_\ell \langle u_j \rangle_\ell  
			\langle u_k \rangle_\ell\,.
\end{align}
Statistical moments of the fluctuations are obtained from the moments of the 
total fields and their averages as, for example,
\begin{align}\label{eq:second_order_moment_direct}
\langle u'_{i} u'_{j} \rangle_\ell &= 
	\left\langle(u_i-\langle u_i\rangle_\ell) 
			(u_j-\langle u_j\rangle_\ell)\right\rangle_\ell %\notag \\
%&
= \left\langle u_i u_j-\langle u_i\rangle_\ell u_j-u_i\langle u_j \rangle_\ell  
	+\langle u_i\rangle_\ell \langle u_j \rangle_\ell\right \rangle_\ell\notag \\
&= \left\langle u_i u_j - \langle u_i\rangle_\ell u'_j-u'_i\langle u_j \rangle_\ell
	- \langle u_i\rangle_\ell \langle u_j\rangle_\ell\right\rangle_\ell\notag \\
&= \langle u_i u_j \rangle_\ell
	- \left\langle\langle u_i\rangle_\ell u'_j \right\rangle_\ell
		- \left\langle u'_i \langle u_j\rangle_\ell\right\rangle_\ell
		-\left\langle\langle u_i\rangle_\ell\langle u_j\rangle_\ell\right \rangle_\ell\,.
\end{align}
As in equation~\eqref{eq:conditions_filtering}, we have
$\left\langle \langle u_i\rangle_\ell u'_j \right\rangle_\ell \neq 0$
and 
$\left\langle u'_i\langle u_j\rangle_\ell \right\rangle_\ell \neq 0$.
In addition, 
$\left\langle\langle u_i \rangle_\ell \langle u_j \rangle_\ell\right\rangle_\ell
\neq \langle u_i\rangle_\ell \langle u_j \rangle_\ell$ since
$\left\langle \langle u_i \rangle_\ell \right\rangle_\ell \neq \langle u_i \rangle_\ell$.
As a consequence, 
$\langle u'_i u'_j\rangle_\ell\neq\langle u_iu_j\rangle_\ell 
- \langle u_i\rangle_\ell \langle u_j\rangle_\ell= \mu(u_i,u_j)$.
Replacing statistical moments of the fluctuations such as 
$\langle u'_i u'_j\rangle_\ell$ wherever they appear with generalised
central moments such as $\mu(u_i, u_j)$, leads to governing equations for the 
fluctuations in a mathematically simple form practically identical to that 
obtained under ensemble averaging \citep[see][for the case of MHD 
equations]{Aluie:2017}.  
The algebraic structure of the closure is the same, regardless of the choice
of the filter $\mathcal {G}$. Such a property is called the averaging
invariance of the turbulent equations \citep[see][]{Germano:1992}.

%---------------------------------------------------------------
\label{sect:smoothing}
In application to the ISM simulations,
%\sout{described in Section~\ref{sect:model_summary}}, 
we consider the decomposition of the physical 
fields into mean and fluctuating components with the mean fields obtained via 
filtering with a Gaussian kernel,
\begin{equation}\label{eq:gauss}
%\langle f(\vect{x}) \rangle_{\ell} 
%= \int_{V} f(\vect{x}')G_{\ell}
%(\vect{x}-\vect{x}') \, \mathrm{d}^{3}\vect{x}', \notag \\
G_\ell(\vect{x}) = (2 \pi \ell^2)^{-3/2}\exp\left[-\vect{x}^2/(2\ell^2)\right],
\end{equation}
where $\ell$ is the smoothing length.
We perform this analysis for magnetic field $\vect{B}$, gas density $\rho$
and velocity $\vect{u}$. All averages are denoted with the subscript $\ell$ and
fluctuations with the prime, with the 
exception of magnetic field fluctuations denoted $\vect{b}$:
\begin{align}
\vect{B} &= \vect{B}_\ell + \vect{b}\,, 
&\vect{B}_{\ell} &= \langle \vect{B} \rangle_\ell\,, 
&\vect{b} &= \vect{B} - \vect{B}_\ell\,, \nonumber\\
\rho &= \rho_\ell + \rho'\,, 
&\rho_\ell &= \langle \rho \rangle_\ell\,, 
&\rho'& = \rho - \rho_\ell\,, \nonumber \\
\vect{u} &= \vect{u}_\ell + \vect{u}'\,, 
&\vect{u}_\ell &= \langle \vect{u} \rangle_\ell\,, 
&\vect{u}' &= \vect{u} - \vect{u}_\ell\,.
\label{eq:decomp} 
\end{align}

%-------------------------------------------------------------------------------

\section{The smoothing scale and Fourier spectra}\label{sect:power_spectra}

%-------------------------------------------------------------------------------

The challenge in applying the filtering approach in our context is to 
determine an appropriate smoothing length $\ell$ or its admissible range. We 
note that the mean and fluctuating parts of different variables, e.g., 
$\vect{B}$, $\rho$ and $\vect{u}$, can have different spatial properties and, 
hence, different smoothing lengths may be required to separate the 
fluctuations in different variables. For example, \citet{Hollins:2017} find 
that the correlation lengths of the three variables are different in the 
simulations discussed here. Unlike applications to subgrid turbulence models, 
where $\ell$ is identified with the spatial resolution of a simulation, the 
choice of $\ell$ in the present context is motivated by physical considerations.
Following \citet{Gent:2013b}, we select $\ell$ using the spectral structure of 
each variable as discussed below.

Scale separation between the mean and fluctuation fields is required neither by 
theory based on ensemble averages nor by the filtering technique. Nevertheless, 
it is natural to expect some difference in scales between the two. For example, 
the scale of the mean field in a turbulent dynamo is controlled by deviations 
of the random flow from mirror symmetry and mean velocity shear, whereas 
turbulent scales depend on the nature of the driving forces. Given the 
fundamental difference between the two groups of physical effects, it is 
unlikely that the two parts of magnetic field have similar scales. Since 
deviations from mirror symmetry are usually weak, the scale of the mean field 
is expected to be correspondingly large and to exceed the turbulent scale. 
Arguments of this kind are used to justify the two-scale approach in 
mean-field magnetohydrodynamics 
\citep{Moffatt:1978,Krause:1980,Zeldovich:1983}. However, numerical 
simulations of dynamo systems (including those discussed here) are performed in 
domains that are only moderately larger than the integral scale of the 
simulated random flow \citep[][and references therein]{Brandenburg:2005} which 
precludes any strong scale separation between the simulated mean and fluctuating 
fields. Nevertheless, evidence for such separation is 
usually sought, in the form of a pronounced minimum in the Fourier spectra 
at a scale exceeding the  presumed integral scale of the fluctuations (often, 
the scale at which the random flow is driven by an explicit force) and the 
domain size. In application to the magnetic field, \citet{Gent:2013b} demonstrate 
that the situation can be more subtle and, despite a pronounced difference of 
the two scales (by a factor of two), the Fourier spectrum of the total magnetic 
field may not have a noticeable minimum between them.

%-------------------------------------------------------------------------------
\begin{table*}[tb]
\centering
\caption{\label{tab:Table1}Notation for the total (T), mean (M) and fluctuating 
	(F) fields and their respective Fourier spectra, arising from the filtering decomposition.  
	See section~\ref{sect:power_spectra} for definitions.}
\begin{small} 
\begin{tabular}{l ccc c ccc c ccc c ccc}%
\hline
&&&& \multicolumn{3}{c}{Spectrum} 
& & \multicolumn{3}{c}{Energy density}  & & \multicolumn{3}{c}{Energy}\\
\cline{5-7}	\cline{9-11}	\cline{13-15}\noalign{\vskip 3pt}
&T &M &F   	&T &M &F	&&T &M &F    	&&T &M &F \\ 
\hline
Magnetic field 
&$\vect{B}$ & $\vect{B}_\ell$ & $\vect{b}$ 
&$S_B(k)$ & $S_{B_\ell}(k)$ & $S_b(k)$ & 
&$\langle e_B\rangle_\ell$ & $e_{B_\ell}$ & $e_b$ & 
&${\cal E}_B$ & ${\cal E}_{B_\ell}$ & ${\cal E}_b$ \\
Gas density 
& $\rho$ & $\rho_\ell$ & $\rho'$   
&$S_\rho(k)$ & $S_{\rho_\ell}(k)$ & $S_{\rho'}(k)$ & 
&$\text{---}$ & $\text{---}$ & $\text{---}$ & 
&$\text{---}$ & $\text{---}$ & $\text{---}$ \\
Gas velocity 
& $\vect{u}$ & $\vect{u}_\ell$ & $\vect{u}'$   
&$S_u(k)$ & $S_{u_\ell}(k)$ & $S_{u'}(k)$ & 
&$\langle e_{\rm k}\rangle_\ell$ & $e_{\rm s}$ & $e_{\rm st}, \, e_{\rm t}$ & 
&${\cal E}_{\rm k}$ & ${\cal E}_{\rm s}$ & ${\cal E}_{\rm st}, \, {\cal E}_{\rm t}$ \\
\hline
\end{tabular}
\end{small}
\end{table*}
%-------------------------------------------------------------------------------

The Fourier spectrum of the total magnetic field $\vect{B}$ is given by 
\begin{equation}\label{eq:S(k)}
S_B(k)=k^2\langle|\widehat{\vect{B}}(\vect{k})|^2\rangle_{k}\,,
\end{equation}
where $\widehat{\vect{B}}(\vect{k})=
\int_V \vect{B}(\vect{x})\exp(-2\pi\ii\vect{k}\cdot\vect{x})\,\dd^3\vect{x}$
is the Fourier transform of $\vect{B}$ and $\langle\cdot\rangle_k$ denotes
the average value within a spherical shell of thickness $\delta k$ with radius
$k=|\vect{k}|$. The power spectra for the mean and random fields,
$S_{B_\ell}(k)$ and $S_b(k)$, are similarly defined in terms of
$\widehat{\vect{B}}_\ell(\vect{k})$ and $\widehat{\vect{b}}(\vect{k})$, the
Fourier transforms of $\vect{B}_{\ell}$ and $\vect{b}$:
\begin{equation}\label{eq:S_mean(k)_S_turb(k)}
S_{B_\ell}(k) = k^2\langle| \widehat{\vect{B}_\ell}(\vect{k})|^2\rangle_{k}\,, 
\qquad
S_b(k) = k^2\langle|\widehat{\vect{b}}(\vect{k})|^2\rangle_k\,.
\end{equation}
We also consider the integral scale of each field 
\citep[Sect.~12.1 in][]{Monin2:1975},
\begin{equation}\label{eq:int_scales}
L = \frac{\pi}{2} 
\frac{\int_{2\pi/D}^{\pi/\Delta} k^{-1} S(k)\,\dd k}{\int_{2\pi/D}^{\pi/ 
\Delta} S(k)\,\dd k} \, ,
\end{equation}
calculated using the appropriate power spectrum $S(k)$,
where $\Delta$ is the grid spacing and $D$ the size of the computational domain.
Since both the mean field and the fluctuations are inhomogeneous, 
equation~\eqref{eq:int_scales} can be used to derive the characteristic scales 
of both the mean and fluctuating fields: 
e.g., $L_{B_\ell}$ for the mean magnetic field,
such that $L_{B_\ell}^2 \simeq |\vect{B}_\ell|/|\nabla^{2}\vect{B}_{\ell}|$.
The spectra and lengths scales for $\rho$, $\vect{u}$ and their respective
mean and fluctuations are defined in a similar manner and denoted
$S_\rho(k)$, $S_{\rho_\ell}(k)$, $S_{\rho'}(k)$, 
$S_u(k)$, $S_{u_\ell}(k)$ and $S_{u'}(k)$, with the corresponding length scales 
$L_\rho$, $L_{\rho_\ell}$, $L_{\rho'}$, $L_u$, $L_{u_\ell}$ and $L_{u'}$. The 
notation is summarized in Table~\ref{tab:Table1}.

%%-------------------------------------------------------------------------------

As discussed in Sections~\ref{sect:magnetic_field_ps}, \ref{sect:density_ps}
and \ref{sect:velocity_ps}, none of the Fourier spectra of $\vect{B}$, $\rho$
and $\vect{u}$, have a local minimum.  Nonetheless, each variable has
distinct, well separated length scales for the mean and fluctuating fields.
The optimal smoothing scale $\ell$ for each variable is obtained 
under the requirements that: (i)~the major
maxima in the Fourier spectra of the mean fields and fluctuations in each
variable occur on different sides, along the wavenumber axis, of the
wavenumber where they intersect; and (ii)~that the ratio of the integral
scales of the mean fields and the fluctuations is (approximately) maximized. 

The power spectrum $S_B(k)$ is equivalent, up to a constant factor of 
$1/(8\pi)$, to the magnetic energy spectrum $M(k)=S_B(k)/(8\pi)$, and
the total magnetic energy can be obtained as an integral over the 
relevant wavenumber range, $E_B=\int_{k} M(k)\,\dd k$. However, unlike the 
case of incompressible flows, the power spectrum of the velocity field cannot 
be directly equated to the kinetic energy density because of the contribution 
from the gas density fluctuations.

Calculation of energy densities due to the mean fields and fluctuations should 
be done with care. To illustrate the general approach, consider magnetic 
energy. 
In the filtering approach, the energy densities sum as required from 
equation~\eqref{eq:generalised_moments_equations}, with the following definitions:
\begin{equation}
\langle e_{B} \rangle_{\ell} = e_{B_\ell} + e_b,\qquad
e_{B}=B^2/(8\pi), \qquad 
e_{B_{\ell}}=B_{\ell}^2/(8 \pi), \qquad
e_b=\mu(B_i,B_i)/(8\pi)
\end{equation}
(with summation over repeated indices understood in the final definition).
Note that $e_b\neq b^2/(8\pi)$; this is discussed further below.
%the `naive' squares of the decomposed quantities do not meaningfully sum within this approach.
%
We introduce a distinct notation for the volume integrals of these energy densities,
to allow the meaningful summation of the energies:
\begin{equation}
{\cal E}_{B}={\cal E}_{B_{\ell}}+{\cal E}_b, \qquad
{\cal E}_{B} = \int_{V} \langle e_{B} \rangle_{\ell} \, {\rm d}V,\qquad
{\cal E}_{B_{\ell}} = \int_{V} e_{B_{\ell}} \, {\rm d}V, \qquad
{\cal E}_{b} = \int_{V} e_{b} \, {\rm d}V,
\end{equation}
where $\dd V=\dd^{3}\vect{x}$.
These energy densities and their volume integrals are summarised 
in Table~\ref{tab:Table1}, 
and discussed further in section~\ref{sect:energy_densities}.

It is important to appreciate the distinction, within the filtering approach,
between the quantities introduced above (which allow a meaningful decomposition of energy),
and the `naive' energies obtained directly from the decomposed parts of the field
(which differ in some cases, and which do not meaningfully sum).
Although energies can be obtained from the latter ---
via volume integrals of the squared quantities,
or via wavenumber integrals of the power spectra ---
these quantities do not sum to a valid decomposition,
and so are not here identified with the energies 
of the mean and fluctuating parts, or of their sum.

For clarity, we do briefly discuss these alternative definitions of energies here, 
but we do not include them in Table~\ref{tab:Table1},
or use these definitions in the following.
The total magnetic energy $E_B$ satisfies
$E_B=1/(8\pi) \int_k S_B(k)\,\dd k=\int_{V} e_B \, \dd V$
(where $e_{B}=B^2/(8\pi)$, as above);
but $E_B\neq{\cal E}_{B}$.
The integral of the mean field satisfies
$E_{B_\ell}=1/(8\pi) \int S_{B_\ell}(k)\,\dd k = \int_V e_{B_\ell} \, \dd V$
(where $e_{B_{\ell}}=B_{\ell}^2/(8 \pi)$, as above),
and does equate to the mean energy introduced above,
$E_{B_\ell} = {\cal E}_{B_\ell}$.
But the corresponding quantities for the fluctuating field,
$E_{b}=1/(8\pi) \int S_b(k)\,\dd k = 1/(8\pi) \int_{V} b^2 \, \dd V$,
and ${\cal E}_{b}= \int_{V} e_b \, \dd V$,
are not equal: $E_{b}\neq{\cal E}_{b}$.
As noted above, and discussed in more detail in section~\ref{sect:energy_densities},
$e_{b}$ must be defined in terms of the generalised second moment with $i=j$ 
(from equation~\eqref{eq:generalised_moments_equations}),
$e_b=\mu(B_i,B_i)/(8\pi)$,
so that $e_b\neq b^2/(8\pi)$.

To summarise the comparison of energies between the two approaches:
${\cal E}_{B_{\ell}}=E_{B_{\ell}}$,
but ${\cal E}_{B}\neq E_{B}$, and ${\cal E}_{b}\neq E_{b}$;
and as noted before, 
while ${\cal E}_B={\cal E}_{B_\ell}+{\cal E}_b$,
$E_B\neq E_{B_\ell}+E_b$.
As a result, we focus in the following on the filtered energies (${\cal E}_{B}$, etc.),
and do not further refer to the naive energies ($E_B$, etc.).
We do analyse the Fourier spectra of the basic physical variables in 
Sections~\ref{sect:magnetic_field_ps}--\ref{sect:velocity_ps} to identify  
appropriate smoothing lengths $\ell$, which can be different for different 
variables; but then use the filtering approach to derive and discuss the 
corresponding energy densities in Section~\ref{sect:energy_densities}.

%-------------------------------------------------------------------------------

\subsection{Magnetic field}\label{sect:magnetic_field_ps}

\begin{figure*}[tb]
\centering
\includegraphics[width=0.46\textwidth]{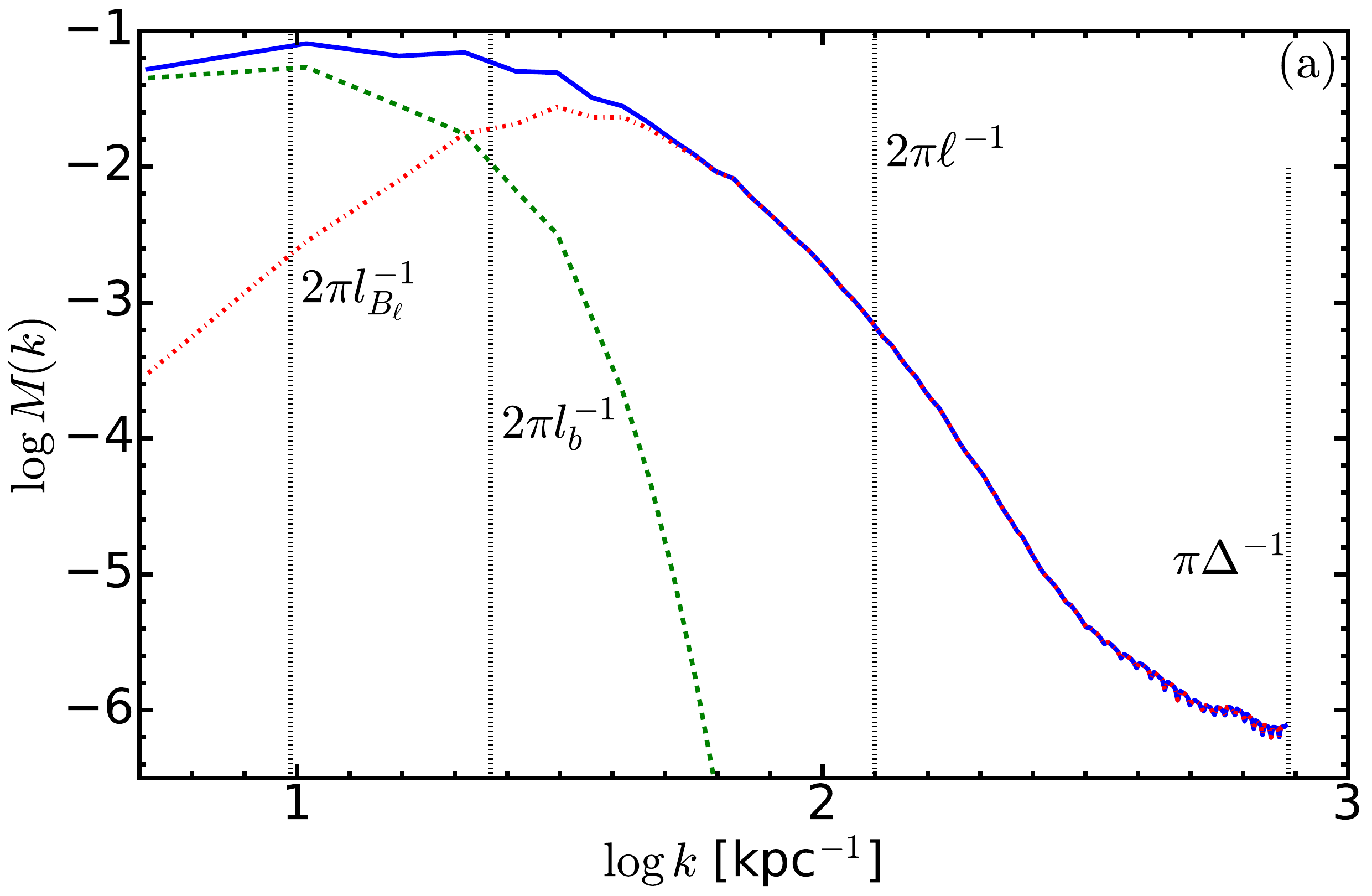}
\hfill
\includegraphics[width=0.46\textwidth]{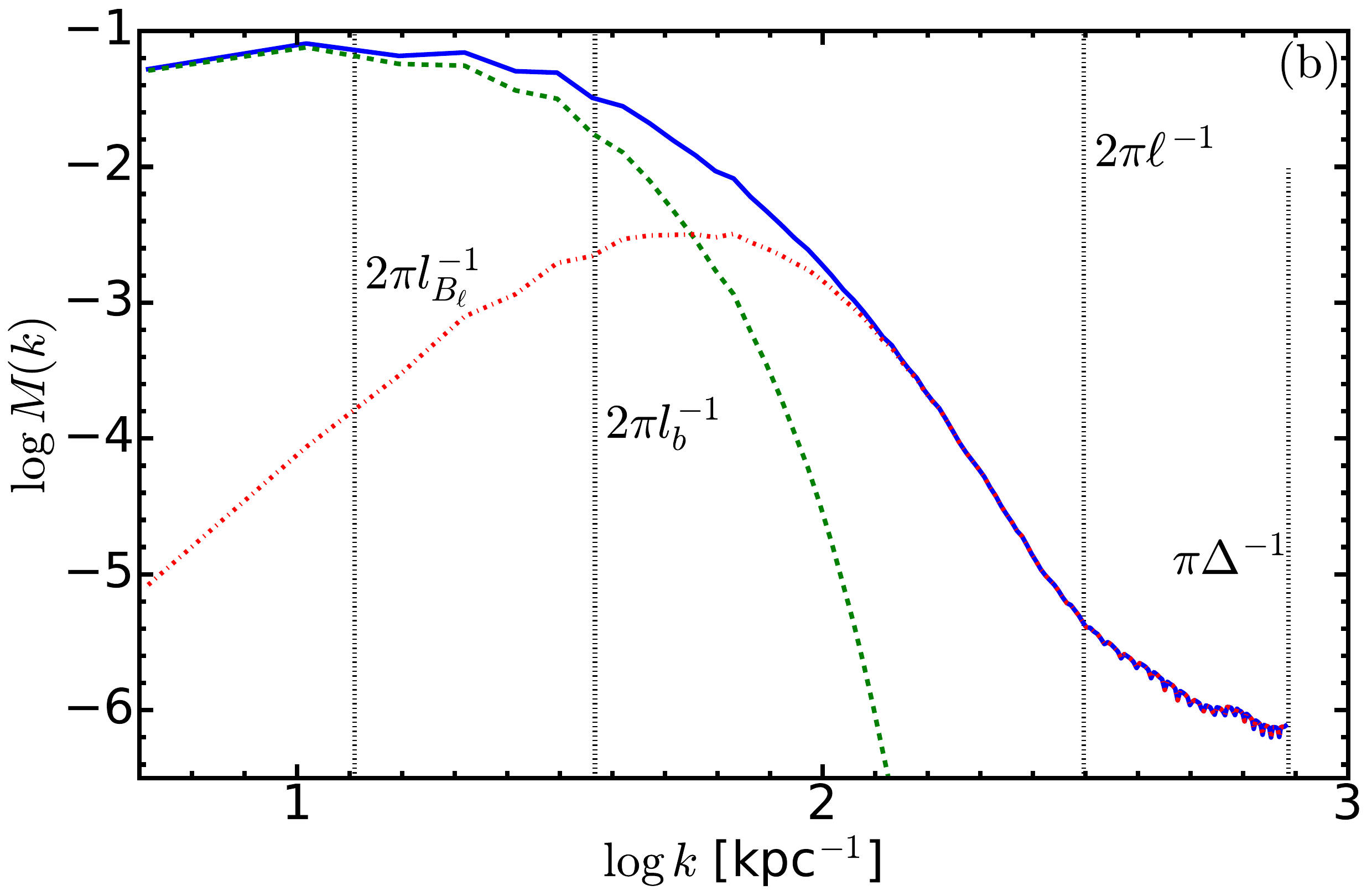}
\\
\includegraphics[width=0.46\textwidth]{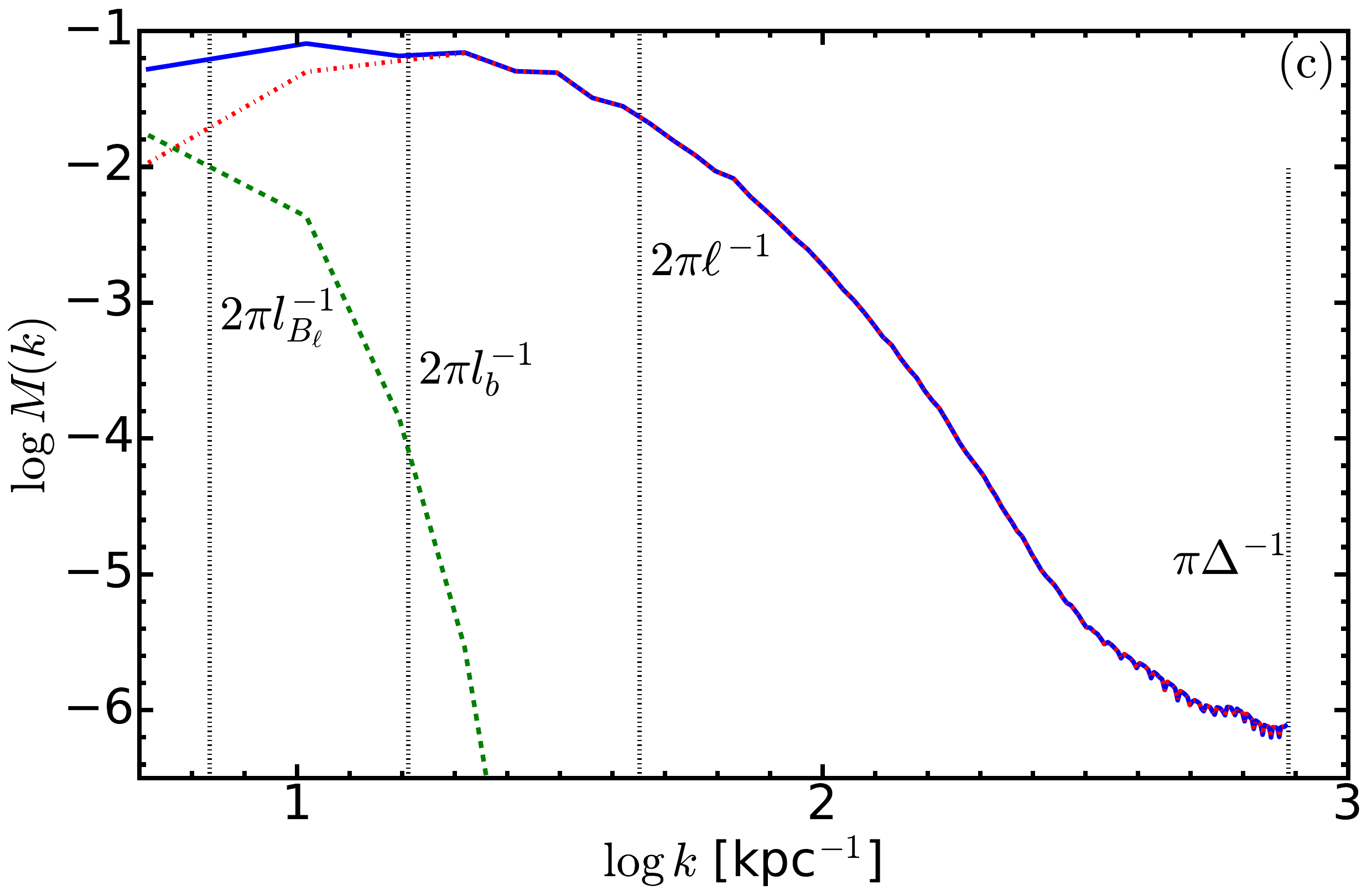}
\hfill
\includegraphics[width=0.46\textwidth]{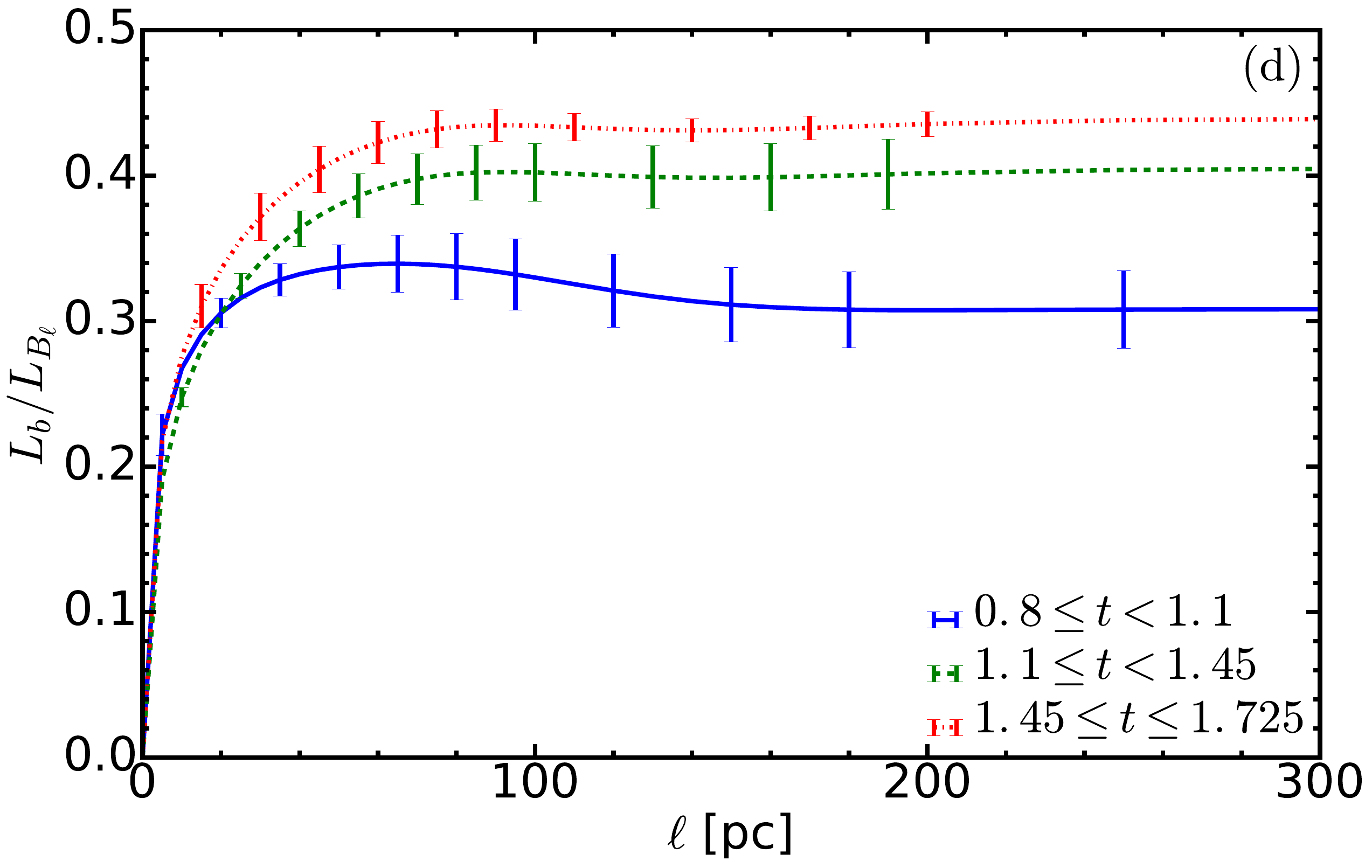}
\caption{\label{fig:b_spec_50_1}Fourier spectra of the total magnetic field 
\label{fig:b_ratio}
$S_B(k)$ (solid, blue), its mean part $S_{B_\ell}(k)$ (dash-dotted, green) and the 
fluctuations $S_b(k)$ (dashed, red) for $\ell=50\p$ at $t=1.6\Gyr$
for various values of the smoothing length $\ell$:
\textbf{(a)}~$\ell=50\p$,
\textbf{(b)}~$\ell=20\p$ and
\textbf{(c)}~$\ell=140\p$. The vertical dotted lines indicate (from left to 
right) the wavenumbers corresponding to the scale of the mean field 
$L_{B_\ell}$, its fluctuations $L_b$, the smoothing length $\ell$ and the 
resolution of the simulations $\Delta$.
\textbf{(d)}: Ratio of the integral scales $L_b$ and $L_{B_\ell}$ 
as a function of the smoothing length $\ell$ in the three stages of magnetic
field evolution, kinematic $0.8 \leq t < 1.1\Gyr$ (solid, blue), transitional 
$1.1 \leq t < 1.45\Gyr$ (dash-dotted, green) and non-linear 
$1.45 \leq t \leq 1.725\Gyr$ (dashed, red).
}
\end{figure*}

In figure~\ref{fig:b_spec_50_1} 
we show the effect of varying the smoothing length $\ell$ on the power
spectra of the mean magnetic field and its fluctuations.
A short $\ell=20\p$ (figure\,\ref{fig:b_spec_50_1}b) assigns the majority of the 
energy to the mean field including a large proportion at small scales,
while too long $\ell=140\p$ (figure\,\ref{fig:b_spec_50_1}c) assigns most of the energy to the fluctuations even at very large scales.
At wavelength $\lambda$ the mean and fluctuation power spectra intersect,
$S_{B_\ell}(\lambda)=S_b(\lambda)$, with $k>\lambda$
mainly characteristic of fluctuations and $k<\lambda$ of mean field.
We also compute the integral scale $L$ for the decomposed fields, and would
expect $L_{B_\ell}>\lambda>L_b$.
Instead for $\ell=20\p$, $\lambda=0.09\kpc<L_b=0.17\kpc$, and for $\ell=140\p$
$L_{B_\ell}=0.92\kpc<\lambda=1.09\kpc$, which are both physically inconsistent.

A more satisfactory picture emerges when $\ell=50\p$, shown in
figure~\ref{fig:b_spec_50_1}a.
$L_{B_\ell}=0.65\kpc$, $\lambda=0.3 \kpc$ and $L_b=0.27\kpc$, such that $L_b<\lambda<L_{B_\ell}$.
Thus, $\ell=50\p$ could be adopted as an appropriate smoothing length for the
magnetic field: then the mean field dominates at scales around $L_{B_\ell}$
whereas the fluctuations contribute most of the power at scales around $L_b$.

%-------------------------------------------------------------------------------
%-------------------------------------------------------------------------------

The ratio of $L_{B_\ell}$ and $l_b$ as a function of $\ell$ is shown in
figure~\ref{fig:b_ratio}d for the three stages of the magnetic field
evolution. When magnetic field is still weak, there is a pronounced maximum at
$\ell=65\p$ which becomes less prominent as the magnetic field growth
saturates. Thus, the requirement that $L_b< \lambda<L_{B_\ell}$ is compatible
with the maximum scale separation between the mean field and the fluctuations.
The ratio reaches an asymptotic value in the range 0.3--0.4 at
$\ell\approx90\p$.

%-------------------------------------------------------------------------------

\subsection{Gas density}\label{sect:density_ps}

%-------------------------------------------------------------------------------
\begin{figure*}[tb]
\centering
\includegraphics[width=0.46\textwidth]{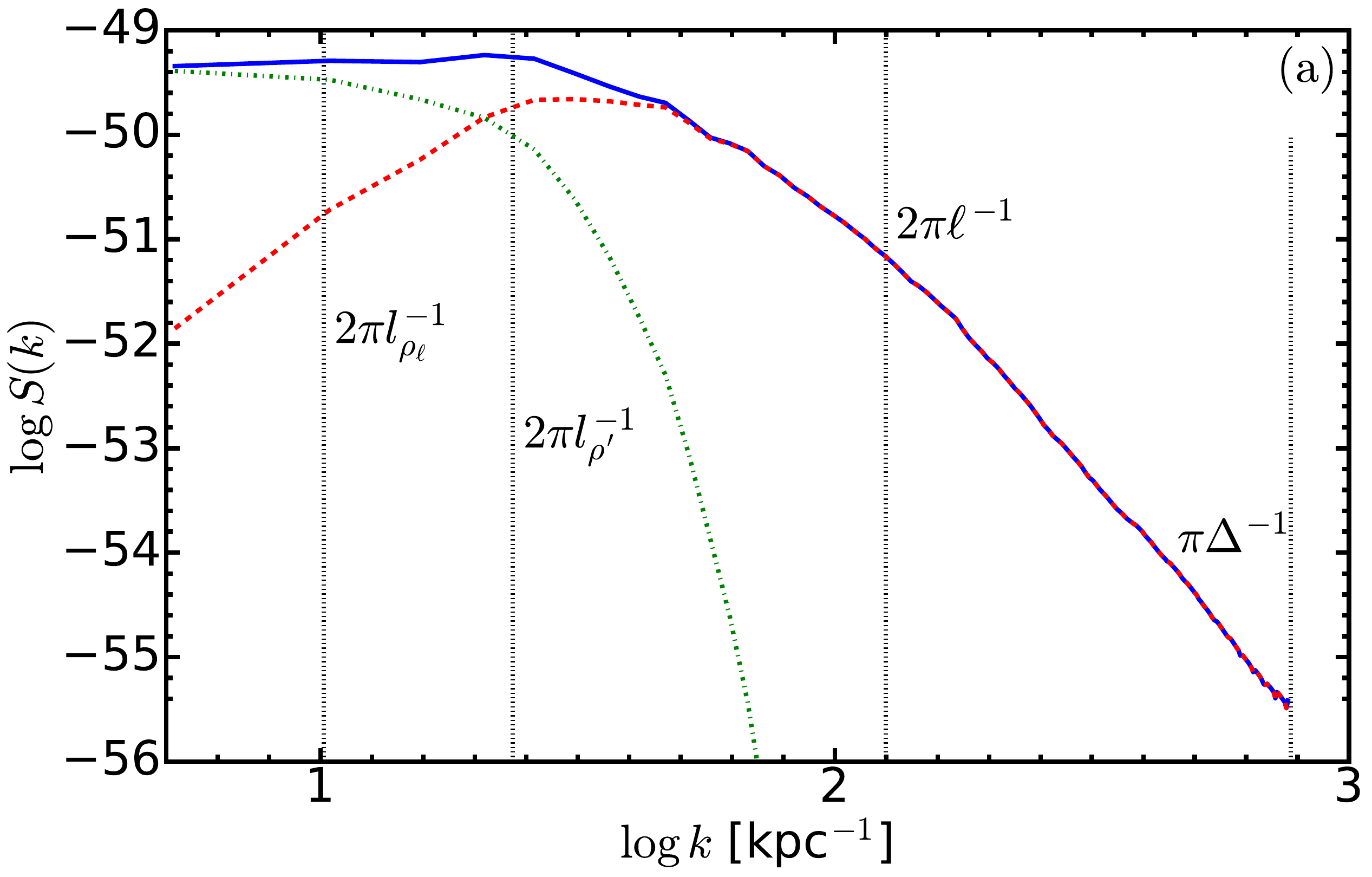}
\hfill
\includegraphics[width=0.46\textwidth]{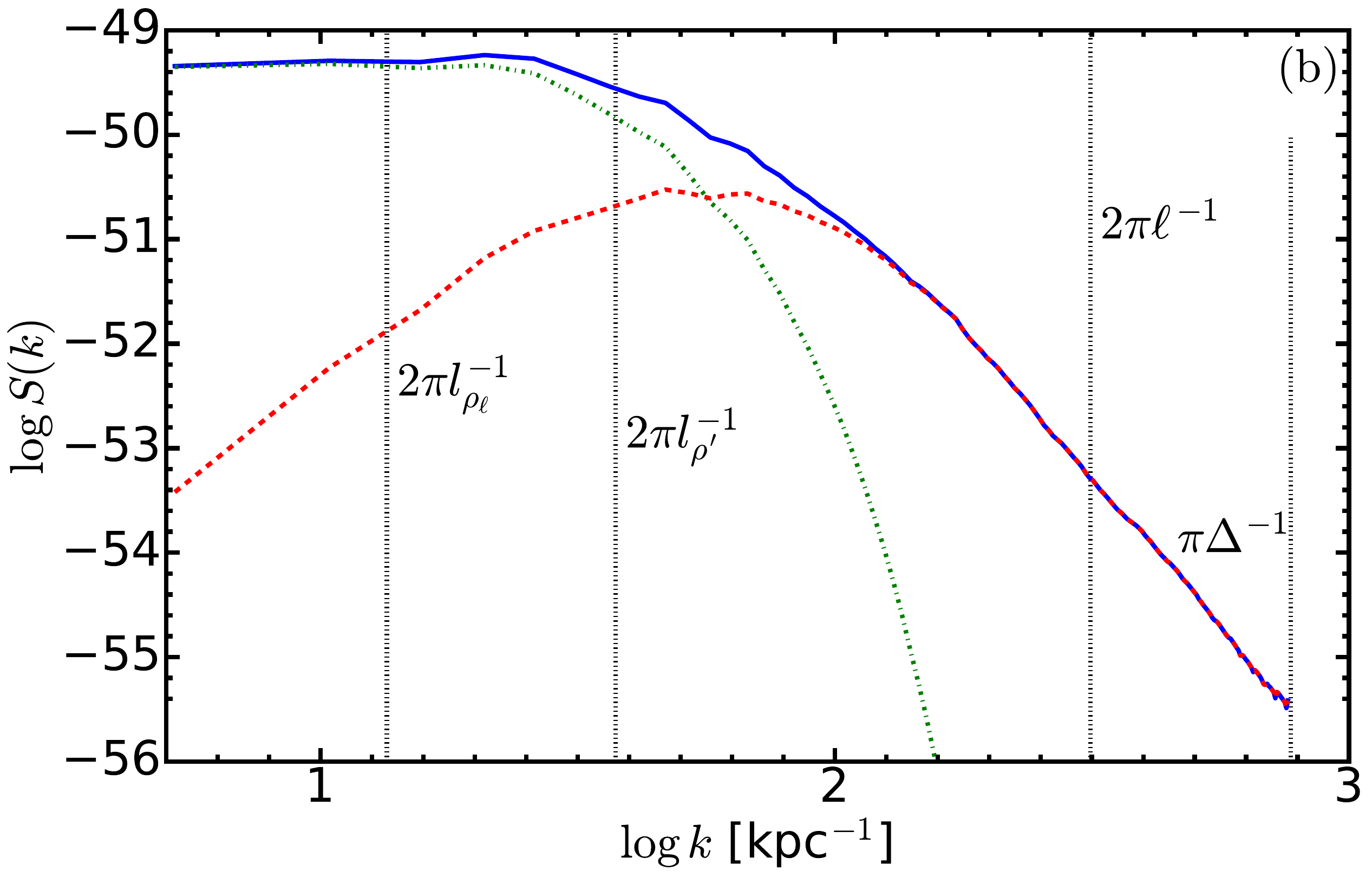}
\\
\includegraphics[width=0.46\textwidth]{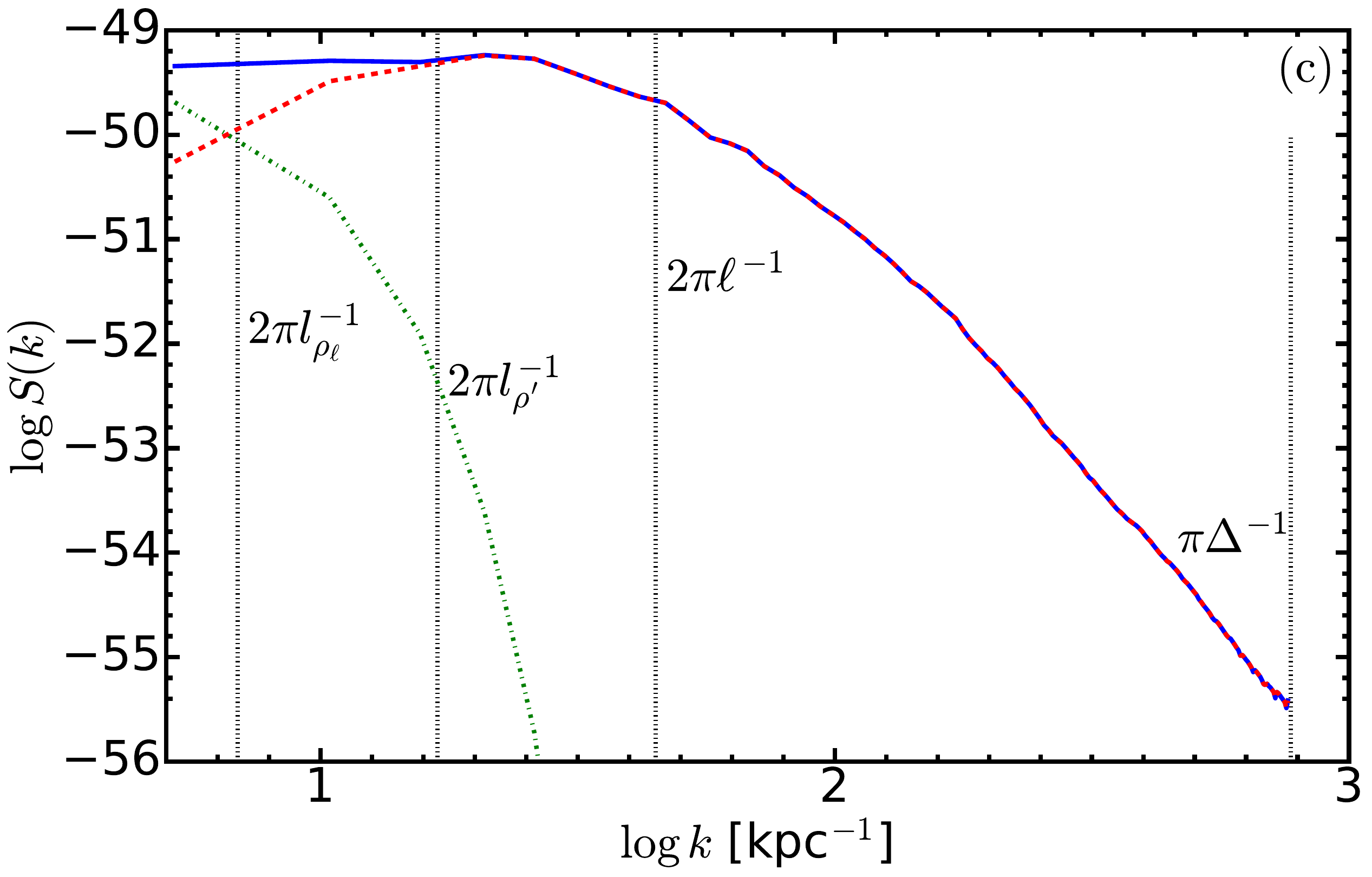}
\hfill
\includegraphics[width=0.46\textwidth]{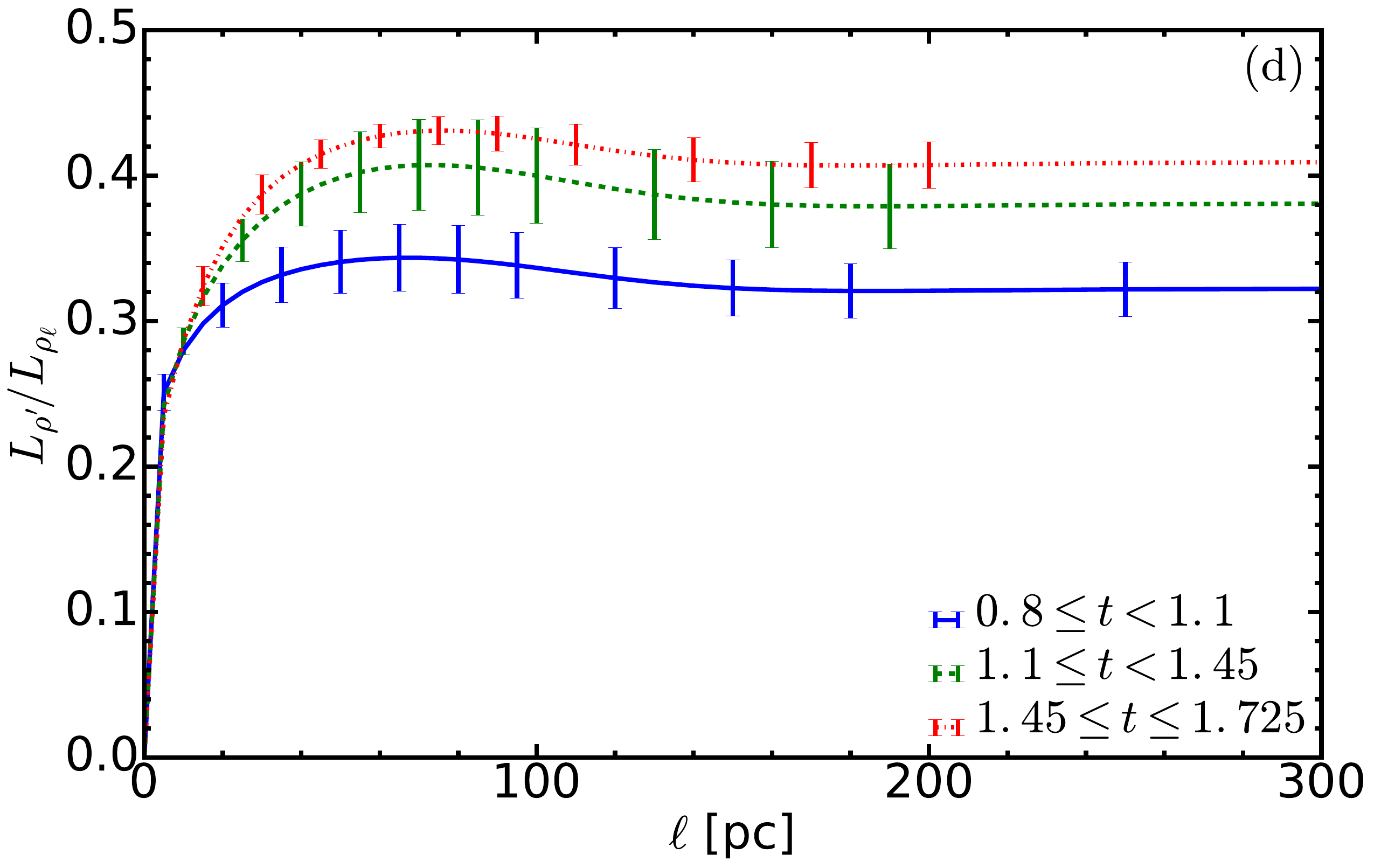}
\caption{\label{fig:n_spec_50}
\label{fig:rho_ratio}
As figure~\ref{fig:b_spec_50_1} but for the gas density $\rho$ (in $\g \cm^{-3}$)
with \textbf{(a)}~$\ell=50\p$, \textbf{(b)}~$\ell=20\p$ and 
\textbf{(c)}~$\ell=140\p$.
}
\end{figure*}
%-------------------------------------------------------------------------------
Using the same arguments as for magnetic field, we conclude that $\ell=50\p$
is a suitable smoothing length for the density distribution, as also shown in
figure~\ref{fig:n_spec_50}. Indeed, when $\ell=50\p$ (figure~\ref{fig:n_spec_50}a), we obtain
$L_{\rho_\ell}=0.62\kpc>\lambda=0.31\kpc>L_{\rho'}=0.27\kpc$.
In contrast for $\ell=20\p$ (figure~\ref{fig:n_spec_50}b), $\lambda=0.11\kpc<L_{\rho'}=0.17\kpc$ and, 
for $\ell=140\p$ (figure~\ref{fig:n_spec_50}c), $L_{\rho_\ell}=0.91\kpc<\lambda=0.95\kpc$.
The ratio of $L_{\rho_\ell}$ and $L_{\rho'}$ as a function of $\ell$ is shown 
in figure~\ref{fig:n_spec_50}d. Its maximum is reached at values of $\ell$ 
increasing from $65\p$ to $75\p$ as the magnetic field saturates, suggesting a 
suitable smoothing length of approximately $70\p$.

%-------------------------------------------------------------------------------

\subsection{Gas velocity}\label{sect:velocity_ps}

%-------------------------------------------------------------------------------
\begin{figure*}[tb]
\centering
\includegraphics[width=0.46\textwidth]{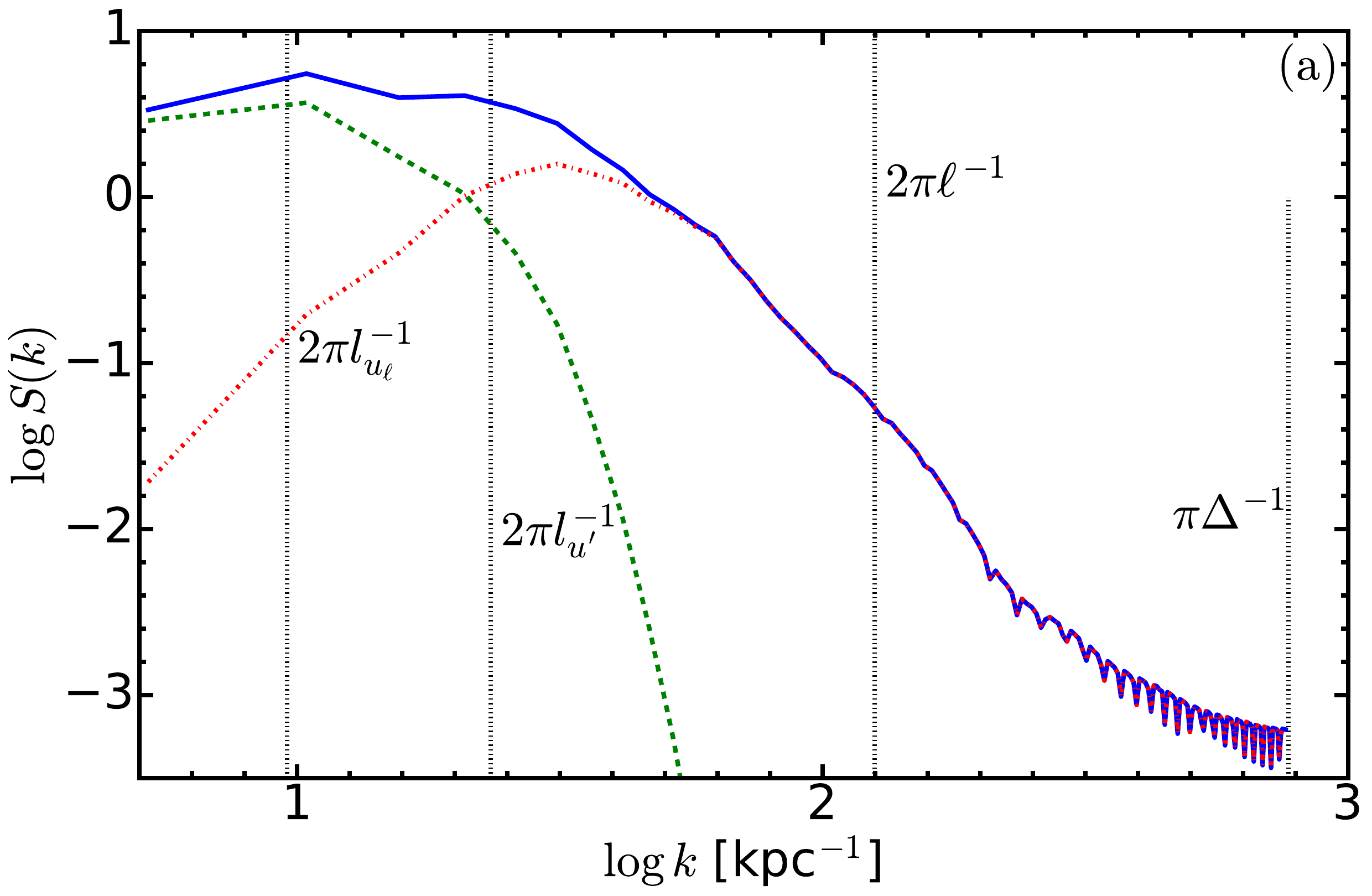}
\hfill
\includegraphics[width=0.46\textwidth]{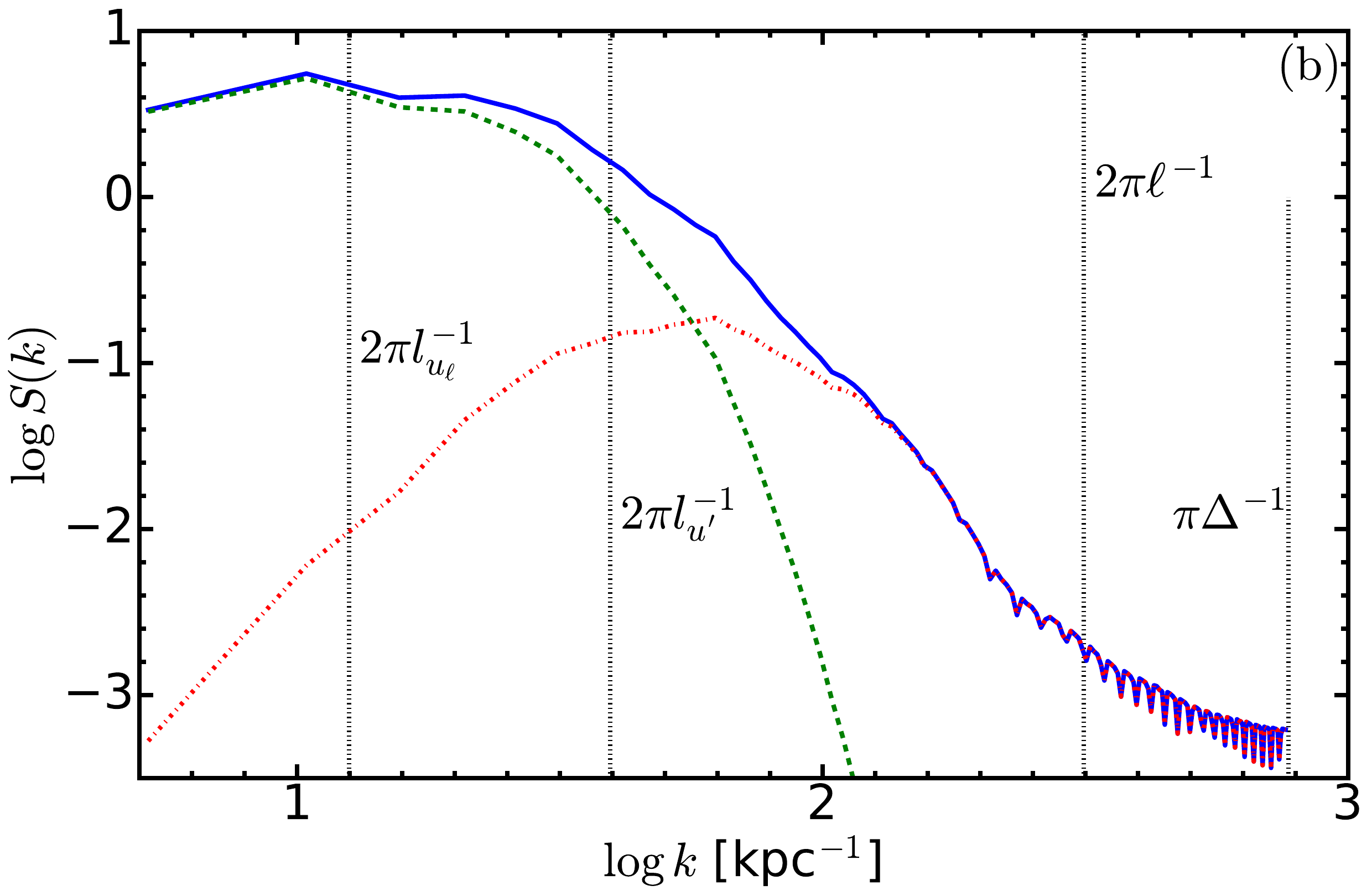}
\\
\includegraphics[width=0.46\textwidth]{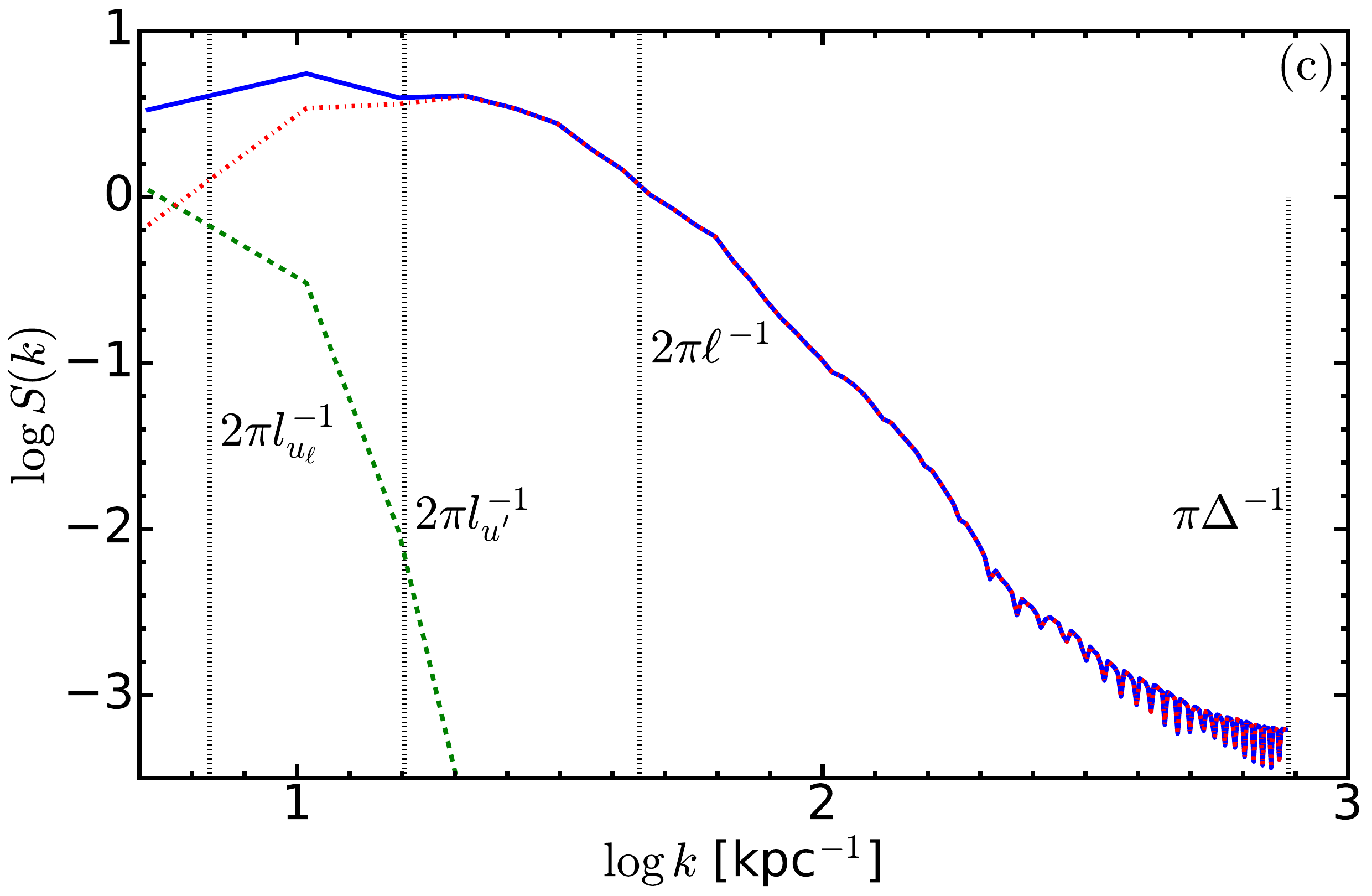}
\hfill
\includegraphics[width=0.46\textwidth]{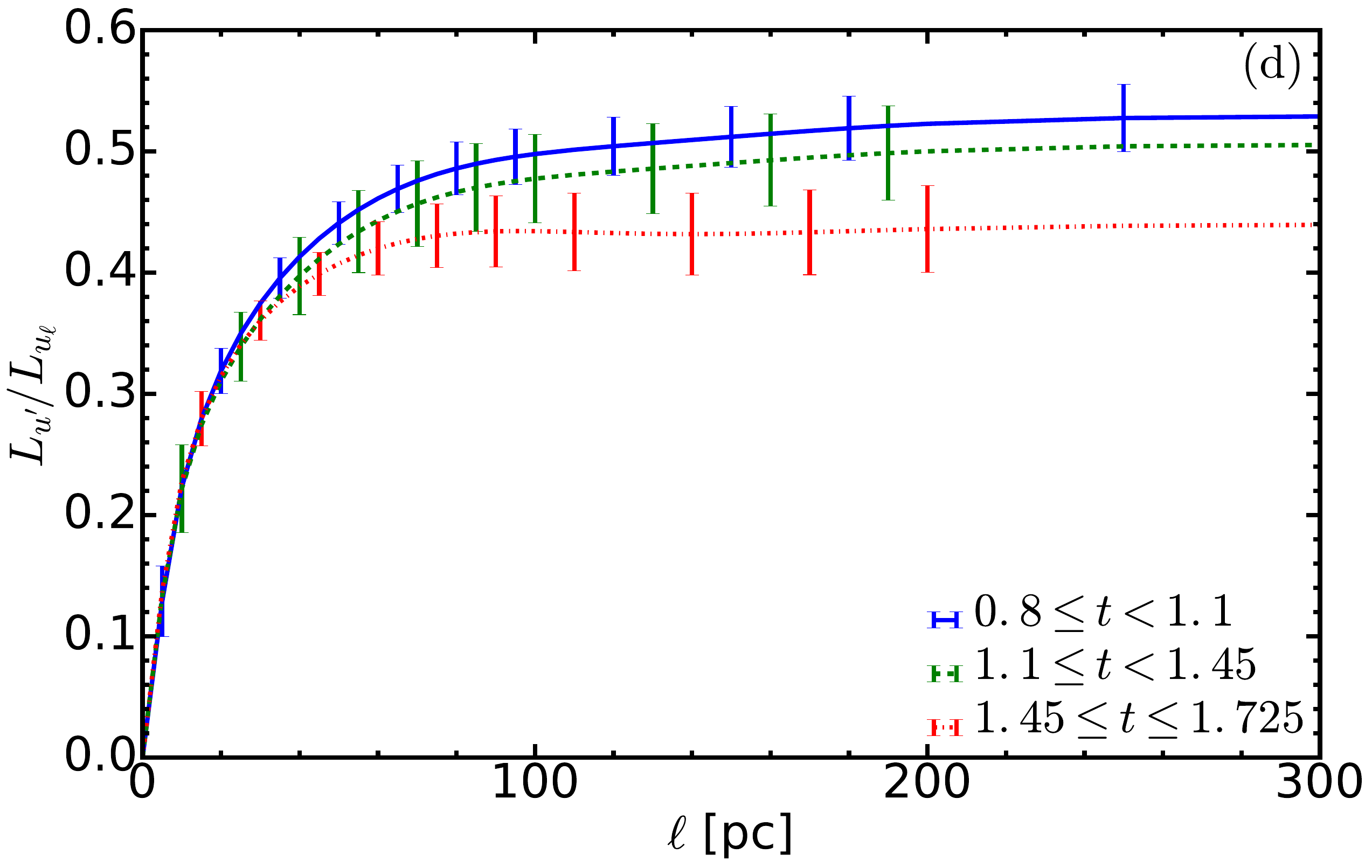}
\caption{\label{fig:u_spec_50}
\label{fig:u_ratio}
As figure~\ref{fig:b_spec_50_1} but for the gas velocity (in $\kms$) with
\textbf{(a)}~$\ell=50\p$, \textbf{(b)}~$\ell=20\p$ and \textbf{(c)}~$\ell=140\p$.
}
\end{figure*}
%-------------------------------------------------------------------------------

Figures~\ref{fig:u_spec_50}a--d illustrate similar arguments for the velocity 
field $\vect{u}$ (we recall that $\vect{u}$ represents deviations from the 
overall shearing flow and contains a systematic vertical outflow velocity). 
When $\ell = 50 \p$, $L_{u_\ell}=0.66\kpc >\lambda=0.3\kpc>L_{u'}=0.27\kpc$.
However, for $\ell = 20 \p$ we have $\lambda=0.12\kpc<L_{u'}=0.16\kpc$, whilst
for $\ell = 140 \p$ we have $L_{u_\ell}=0.92\kpc<\lambda=1.12\kpc$, 
each in conflict with the demarcations of mean and fluctuation field.
Unlike for the magnetic field and gas density spectra, the ratio of length
scales in figure~\ref{fig:u_ratio}d does not have
any pronounced maxima, as it increases monotonically with $\ell$ for $t < 1.45
\Gyr$, and has a very broad maximum at $\ell=90\text{--}100\p$ for $t>1.45
\Gyr$.

It is clear from each of figures~\ref{fig:b_spec_50_1}, \ref{fig:n_spec_50}
and \ref{fig:u_spec_50}, that the spectral properties of each of these fields
are distinct. In addition, the properties of each field vary in time. The 
simulation times considered here, $0.8 \leq t \leq 1.725 \Gyr$, are all well after 
the SN-driven hydrodynamics has reached a statistically-steady state, which occurs at 
about $400 \Myr$. Thus, we are confident that any changes in time result from the 
evolution of the mean-field dynamo, which evolves over a time-scale of order $\Gyr$.

It would therefore seem most appropriate to select different smoothing lengths 
to obtain the fluctuations, depending on both the variable considered and the 
simulation time. However, complications would then arise with the interpretation
of results obtained from such choices. The sensitivity of the results to any change
in smoothing length would have to be considered. 
Theories based on a filtering approach to the MHD equations requires a 
consistent filter as the averaging operator. Hence, applying different smoothing 
lengths for each variable would introduce new difficulties when trying to interpret 
the mean fields and moments of the fluctuating fields as solutions of the filtered 
equations.
In addition, complications could arise when selecting a smoothing scale for moments
computed from multiple basic variables, such as the kinetic energy 
$\rho \boldsymbol{u}^{2}$.
A time dependent smoothing length could be used, interpreted as a change in the grid
scale of such a simulation.

We suggest it most useful to identify an appropriate value of $\ell$
that can be used as a smoothing length for all three variables throughout the
times considered. We adopt $\ell = 75 \p$ as the smoothing length for
magnetic field, gas density and gas velocity, since for magnetic field and gas 
density the local maxima in the ratios of the mean and fluctuating length scales
occur close to $75 \p$. For the gas velocity, the value of this ratio at $75 \p$
is above $90 \%$ of the asymptotic value in each stage, whilst the value at $75 \p$ 
in the saturated stage is very similar to the value at the broad local maximum.

In the context of the numerical model described in 
Section~\ref{sect:model_summary}, this smoothing length 
is comparable to the correlation scales for the density 
fluctuations, random velocity and magnetic fields %in the numerical model, 
as obtained by \citet{Hollins:2017}.
Such a result is sensitive to the choice of simulation 
parameters. For example, an increased supernova rate results
in increased gas compressions, leading to stronger local
density gradients and the formation of more filamentary 
structures. Additionally, stronger local changes in velocity 
would be observed. Thus, the length scales of the small-scale
density and velocity would likely decrease, and so the optimal
choice of $\ell$ would be reduced. Similar effects are likely
to be observed with the reduction of the kinematic viscosity and
the thermal and magnetic diffusivities.

Such a detailed 
exploration of the parameter space, involving multiple simulations
with different parameter choices, is beyond the scope of this 
paper. However, in this section, we have successfully demonstrated
that an optimal choice of the smoothing length for each of the
fundamental physical fields can be reasonably obtained within a
detailed simulation of the ISM.

%-------------------------------------------------------------------------------

\section{Energy densities}\label{sect:energy_densities}

Magnetic and kinetic energy densities have to be derived using the generalized 
central moments, as discussed in Section~\ref{sect:filtering}. The required 
moments are derived in Appendix~\ref{IFSMSTO}. Since the mean 
and fluctuating fields are sensitive to the choice of smoothing length, the 
resultant energies will also depend on $\ell$. The maximum admissible value of 
$\ell$ is half the horizontal extent of the simulation domain. We derive the 
energy densities obtained with various smoothing lengths in the range 
$0<\ell<0.5\kpc$ and discuss the results in this section. 
As previously, we consider the three stages of the mean-field dynamo
separately, and present results averaged over the snapshots within each
stage.

%---------------------------------------------------------------------------
\begin{figure*}[tb]
\centering
\includegraphics[width=0.46\textwidth]{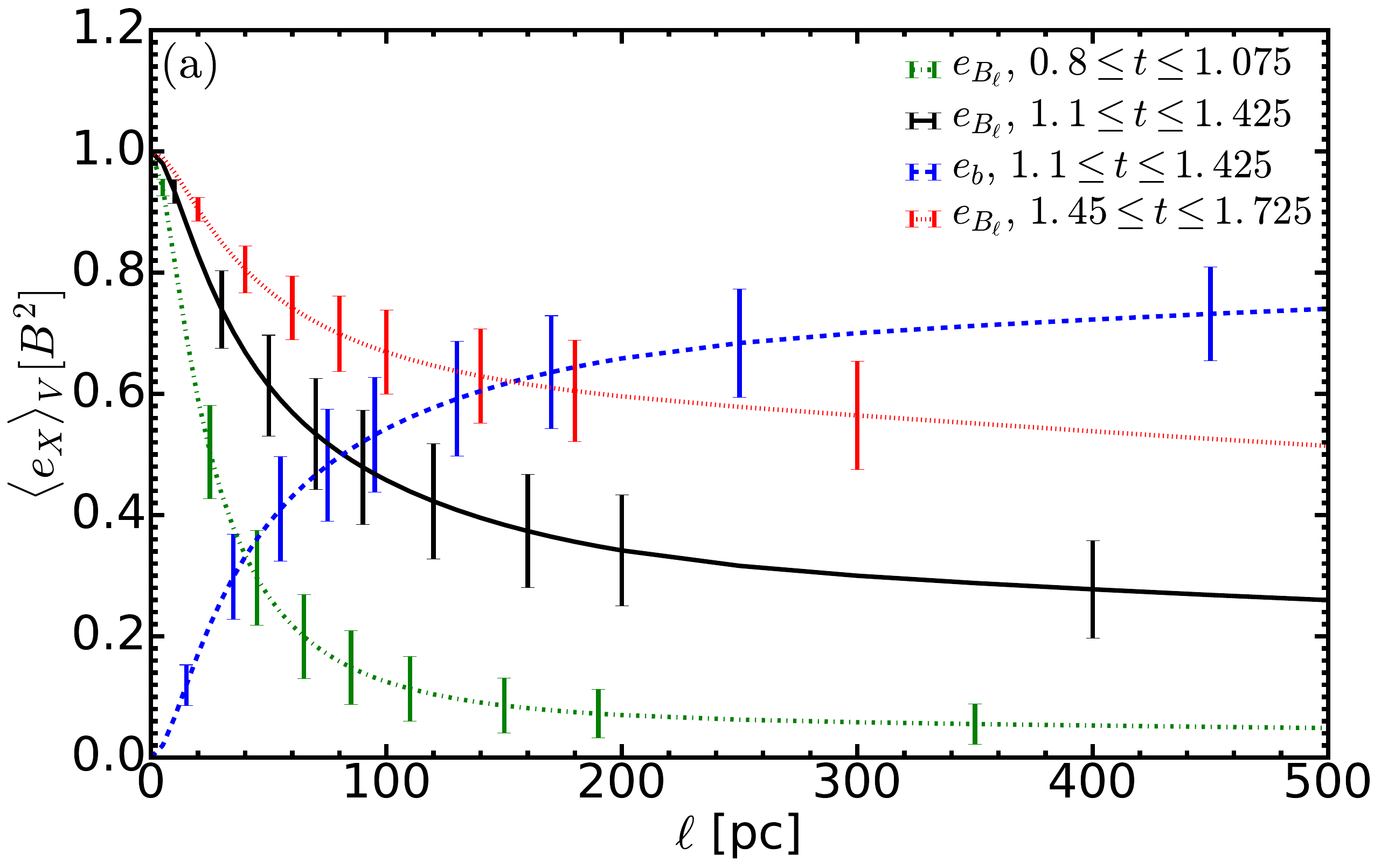}	   
\hfill
\includegraphics[width=0.46\textwidth]{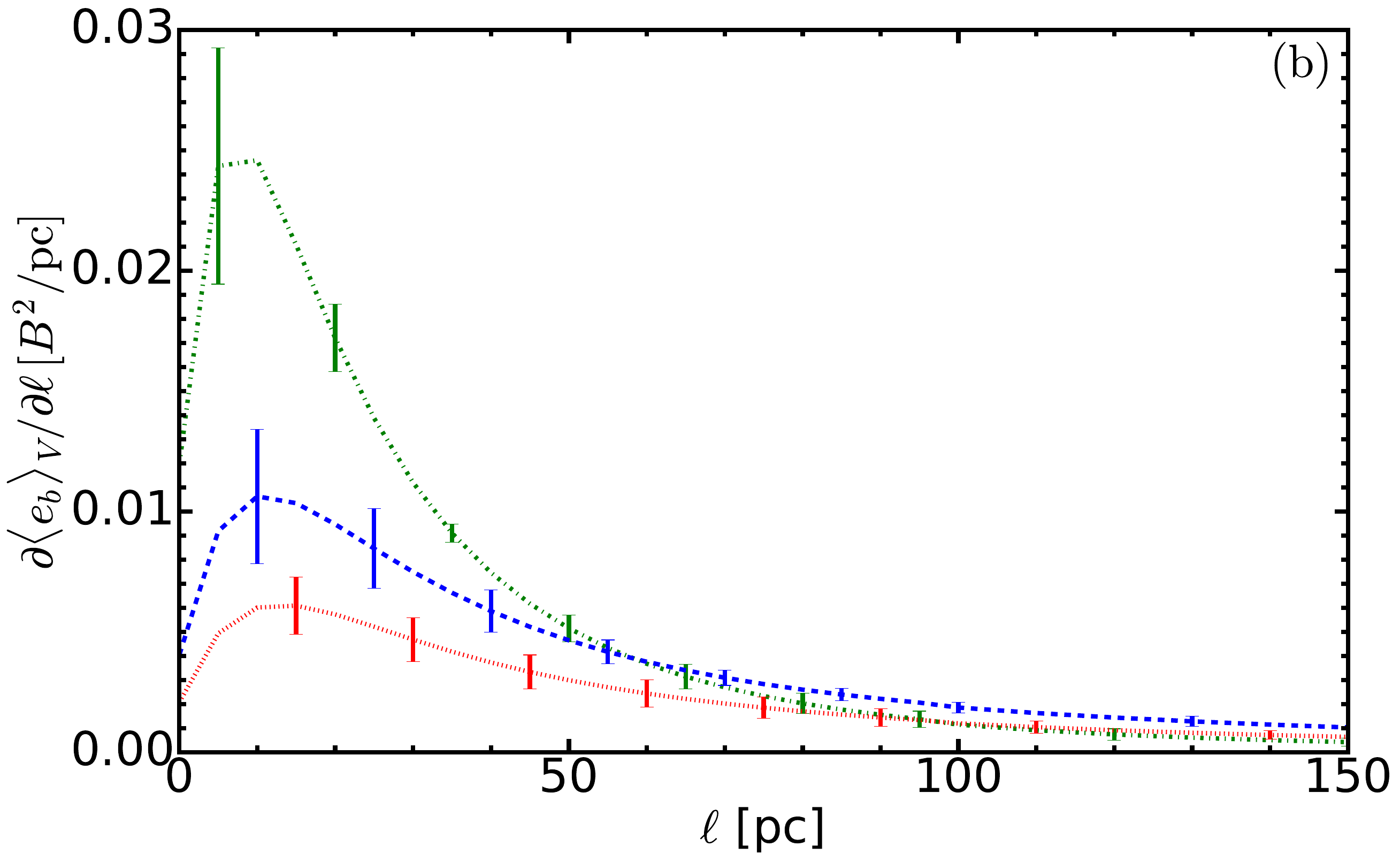}
\caption{a) Volume averages of the mean magnetic energy density 
$\langle e_{B_{\ell}} \rangle_{V}$ at times $0.8 \leq t < 1.1\Gyr$ (green, 
dash-dotted), $1.1 \leq t < 1.45\Gyr$ (black, solid) and $t \geq 1.45\Gyr$ (red, 
dotted); also the fluctuating magnetic energy density $\langle e_{b} \rangle_{V}$ 
at $1.1 \leq t < 1.45\Gyr$ (blue, dashed), as functions of the smoothing length 
$\ell$. These are normalised by the volume average of the smoothed magnetic energy
density, $\langle \langle e_{B} \rangle_{\ell} \rangle_{V}$, with the volume
averaging over the region $|z| < 0.5$.
b) Derivatives of $\langle e_{b} \rangle_{V}$, normalised by 
$\langle \langle e_{B} \rangle_{\ell} \rangle_{V}$, with respect to $\ell$ at
$0.8 \leq t < 1.1\Gyr$ (green,dash- dotted), $1.1 \leq t < 1.45\Gyr$ (blue, dashed) 
and $t \geq 1.45\Gyr$ (red, dotted).
}
\label{fig:e_b}
\end{figure*}
%-------------------------------------------------------------------------------

%-------------------------------------------------------------------------------
\begin{figure*}[tb]
\centering
\includegraphics[width=0.46\textwidth]{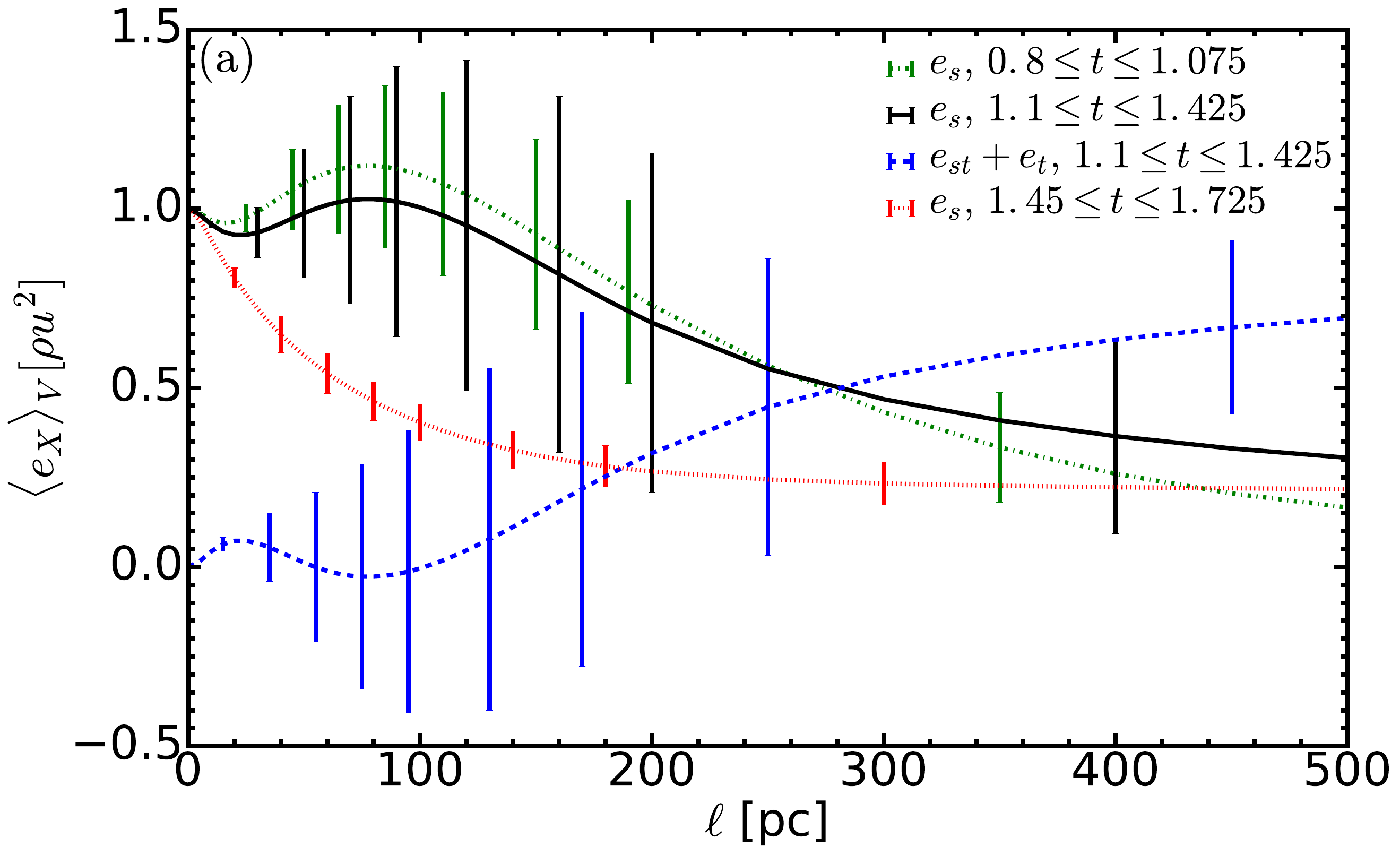}
\hfill
\includegraphics[width=0.46\textwidth]{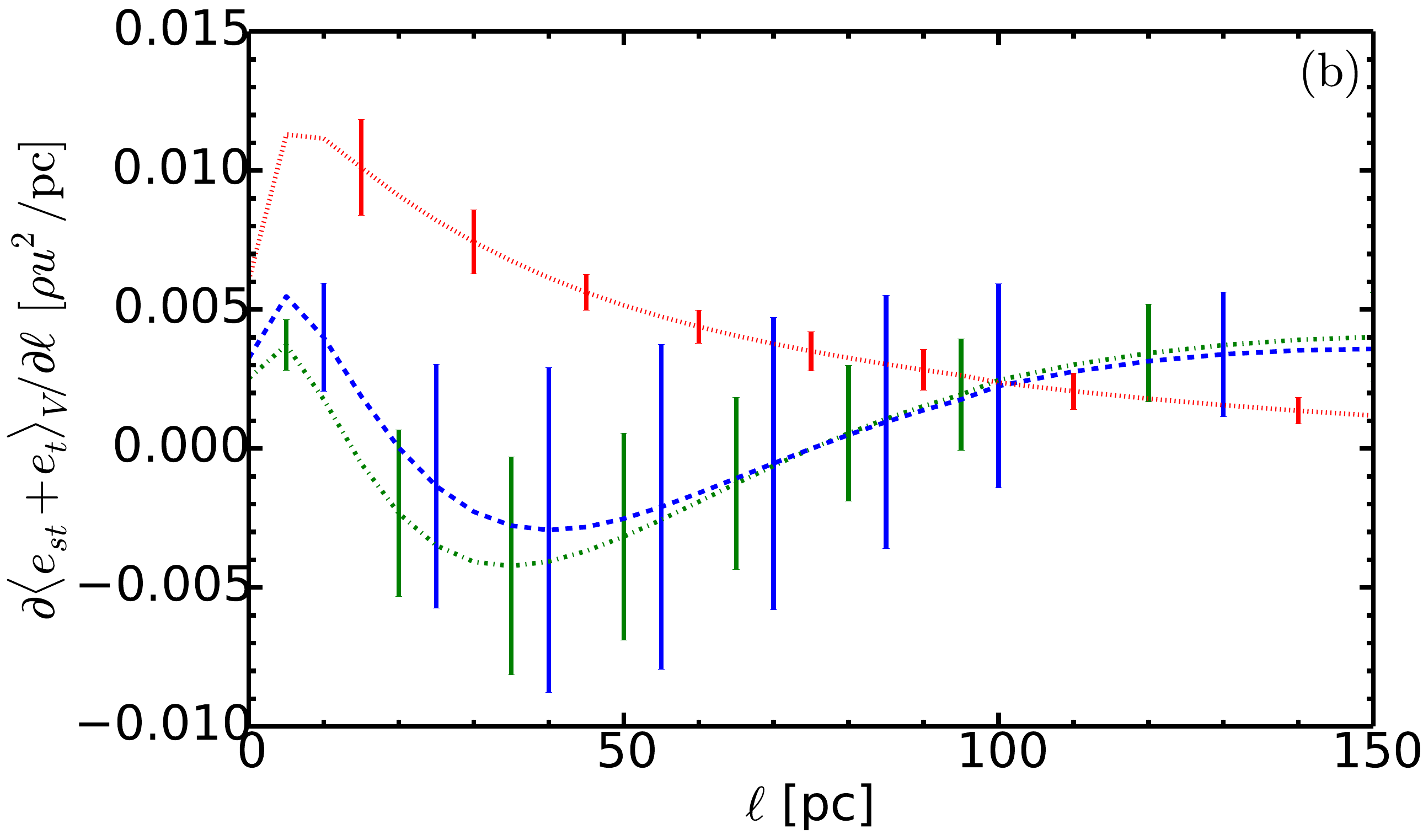}
\caption{\textbf{(a)}~As figure~\ref{fig:e_b}a but for the volume average of the 
mean kinetic energy density $\langle e_\text{s}\rangle_V$ at 
$0.8 \leq t < 1.1\Gyr$ (green, dash-dotted), $1.1 \leq t < 1.45\Gyr$ (black, solid) 
and $t \geq 1.45\Gyr$ (red, dotted); with the volume average of the fluctuating 
kinetic energy density $\langle e_\text{st} + e_\text{t}\rangle_V$ at 
$1.1 \leq t < 1.45\Gyr$ (blue, dashed). These are normalised by the volume average 
of the smoothed kinetic energy $\langle \langle e_\text{k} \rangle_\ell \rangle_V$.
\textbf{(b)}~As figure~\ref{fig:e_b}b but for the derivative of 
$\langle e_\text{st}+e_\text{t}\rangle_V$, with respect to $\ell$ (normalised by 
$\langle \langle e_\text{k} \rangle_\ell\rangle_V$); at $0.8 \leq t < 1.1\Gyr$ 
(green, dash-dotted), $1.1 \leq t < 1.45\Gyr$ (blue, dashed) and $t \geq 1.45\Gyr$ 
(red, dotted).
}
\label{fig:e_u}
\end{figure*}
%-------------------------------------------------------------------------------

%-------------------------------------------------------------------------------

\subsection{Magnetic energy}

The total magnetic energy density is given by
\begin{equation*}
e_B = |\vect{B}|^2/(8\pi)\,,
\end{equation*}
with the energy density of the fluctuating magnetic field obtained as
\begin{equation}
e_{b} = \frac{1}{8 \pi} \int_{V} |\vect{B}(\vect{x}') -
\vect{B_{\ell}}(\vect{x})|^{2} \, 
G_{\ell}(\vect{x}-\vect{x}') \; \mathrm{d}^{3} \vect{x}'\, .
\label{eq:localmag}
\end{equation}
This ensures the energies of the mean and fluctuating magnetic fields sum to the
energy of the (filtered) total magnetic energy, i.e. 
\[
\langle e_{B} \rangle_{\ell} = e_{B_{\ell}} + e_{b} \, ,
\]
where $e_{B_\ell}=|\vect{B}_\ell|^2/(8\pi)$ is the energy density of the mean 
magnetic field. We note that $e_b\neq|\vect{b}|^2/(8\pi)$, but it can be shown, 
by expanding $\vect{B}(\vect{x}')$ in a Taylor series around $\vect{x}$, 
that $e_b=|\vect{b}|^2/(8\pi)+\mathcal{O}(\ell^2/L_{B_\ell}^2)$. 
Thus, the difference between the volume and filtering averages decreases as
$\ell/L_{B_\ell}\to0$. 
This fact, also true for any other variable, suggests one consideration
for the choice of $\ell$ might be to maximise the ratio for $L_{B_\ell}/\ell$.
In practice, however, this would simply lead to $\ell\rightarrow0$; i.e.\
all the signal in the mean field, and effectively no decomposition.

The larger is $\ell$, the smaller part of the total field is deemed to be a 
mean field, and $\langle e_{B_\ell} \rangle_{V}$ monotonically decreases with 
$\ell$ whilst $e_b$ monotonically increases, as shown in figure~\ref{fig:e_b}a. 
The rate of variation of
$\langle e_b\rangle_V/\langle\langle 
e_B\rangle_\ell\rangle_V$ with $\ell$, shown in  figure~\ref{fig:e_b}b
--- and also of $\langle e_B\rangle_V/\langle\langle 
e_B\rangle_\ell\rangle_V$, not shown ---
becomes relatively small when $\ell>50\p$. This confirms that the appropriate choice 
for the smoothing length is $\ell > 50\p$. (The difference between 
figure~\ref{fig:e_b}a and figure~2a of \citet{Gent:2013a} is caused by a
downsampling to a grid $\Delta x=8$\,pc used in the Fourier transform for that 
calculation in \citet{Gent:2013a}.)

The mean magnetic energy grows with time due to dynamo action, and the value of 
$\ell$ for which the two energies are equal to each other increases. At late 
times, the mean magnetic field is energetically dominant over the fluctuating 
magnetic field for all $\ell$.

%-------------------------------------------------------------------------------

\subsection{Kinetic Energy}
In a compressible flow, the mean kinetic energy density is represented by a 
third-order moment involving the density and velocity fields. Under ensemble 
(or volume) averaging, the mean kinetic energy density is conveniently --- and 
physically meaningfully --- represented \citep[see Section~6.4 in][]{Monin:1975} as
\begin{align}\label{eq:compressible_monin_yaglom}
\langle e_\text{k}\rangle& = \tfrac12\langle\rho u_iu_i\rangle\nonumber\\ 
&= \tfrac12\langle\rho\rangle \langle u_i\rangle \langle u_i\rangle 
	+\langle u_i \rangle \langle \rho' u'_i \rangle
		+\tfrac12 \langle \rho u'_i u'_i \rangle \nonumber \\
&\equiv e_{\rm s} + e_{\rm st} + e_{\rm t}\,,
\end{align}
where $e_{\rm s}$ is the energy density of the mean flow, $e_{\rm t}$ is 
the energy density of the fluctuations and $e_{\rm st}$ represents the 
transport of momentum $\langle \rho'u'_i\rangle$ by the mean flow 
(summation over repeated indices is understood here and below). An equivalent
decomposition is appropriate under the filtering approach as well:
\begin{equation}\label{eq:tau_monin_yaglom}
\begin{split}
\langle e_{\rm k}\rangle_\ell&=\tfrac12\langle\rho u_iu_i\rangle_{\ell}
	=e_{\rm s} + e_{\rm st} + e_{\rm t} \,,\\
e_{\rm s} &=\tfrac12\langle\rho\rangle_\ell \langle u_i\rangle_\ell 
		\langle u_i\rangle_\ell\,,\\
e_{\rm st} &= \langle u_i\rangle_\ell\, \mu(\rho, u_i)\,, \\
e_{\rm t} &= \langle e_{\rm k}\rangle_\ell - e_{\rm s} - e_{\rm st}%\\
%&
= \tfrac12\langle\rho\rangle_\ell\, \mu(u_i, u_i)+\tfrac12\mu(\rho,u_i,u_i)\,,
\end{split}
\end{equation}
where the moments involved are derived in Appendix~\ref{IFSMSTO} in explicit 
integral forms:
\begin{align}
e_\text{st} = &\int_V \vect{u}(\vect{x}') G_\ell(\vect{x}-\vect{x}')\, 
			\dd^3 \vect{x}' %\nonumber \\
%	&\mbox{}\quad \times
\int_V \Delta\rho_\ell(\vect{x},\vect{x}')
%	[\rho(\vect{x}') - \rho_\ell(\vect{x})] 
		\Delta\vect{u}_\ell(\vect{x},\vect{x}')
%		[\vect{u}(\vect{x}') - \vect{u_\ell}(\vect{x})] 
					G_\ell(\vect{x}-\vect{x}')\,\dd^3\vect{x}'\,,\nonumber  \\ 
e_\text{t} = &\tfrac12 \int_V \rho(\vect{x}') G_\ell(\vect{x}-\vect{x}')
			\,\dd^{3} \vect{x}'%\nonumber\\
%	&\mbox{\quad}\times 
\int_V |\Delta\vect{u}_\ell(\vect{x},\vect{x}')|^{2}
					G_\ell(\vect{x}-\vect{x}')\,\dd^3 \vect{x'}\nonumber \\\label{eq:localkinetic}
&\mbox{\quad}+\tfrac12\int_V \Delta\rho_\ell(\vect{x},\vect{x}')
%[\rho(\vect{x}') - \rho_\ell(\vect{x})] 
	|\Delta\vect{u}_\ell(\vect{x},\vect{x}')|^2 
			G_\ell(\vect{x}-\vect{x}')\,\dd^3\vect{x}'\,,			
\end{align}
where $\Delta\rho_\ell(\vect{x},\vect{x}')=\rho(\vect{x}') - 
\rho_\ell(\vect{x})$ and  
$\Delta\vect{u}_\ell(\vect{x},\vect{x}')=\vect{u}(\vect{x}') - 
\vect{u_\ell}(\vect{x})$.

Figure~\ref{fig:e_u} shows how various parts of the kinetic energy density 
depend on the smoothing length $\ell$.
The behaviour of the volume averages of these contributions to the kinetic 
energies is much less straightforward than for magnetic energy, except
for $t>1.45 \Gyr$ where similar monotonic dependence on $\ell$ is observed.
Additionally for both $0.8 \leq t < 1.1\Gyr$ and $1.1 \leq t < 1.45\Gyr$, 
we observe that the fluctuating kinetic energy
$\langle e_{\rm st} + e_{\rm t} \rangle_V$
is equal to zero within errors for $50 \leq \ell \leq 100 \p$. 
This results from cancellation between 
$\langle e_\text{st} \rangle_{V}$ and $\langle e_\text{t} \rangle_{V}$,
with $\langle e_\text{st} \rangle_{V}$
significantly negative, as confirmed by figure~\ref{fig:kin_eng_vertical_profiles}.
The quantity $e_\text{st}=\langle u_i \rangle_\ell\, \mu(\rho,u_i)$ is dominated
by the contribution of the $z$-component of the velocity field ($i=3$) since 
$\langle u_z \rangle_{\ell}$ is much larger than the $x$- and $y$- components
because of a systematic gas outflow from the midplane. 

The gas involved in the outflow is hotter and less dense than on average, 
leading to large negative values of 
$-\langle\rho\rangle_\ell \langle u_z\rangle_\ell$ for $z>0$
and, hence, of $\langle u_z\rangle_\ell \, \mu(\rho,u_z)=\langle u_z\rangle_\ell 
\, (\langle\rho u_z\rangle_\ell-\langle\rho\rangle_\ell \langle u_z\rangle_\ell)$
(the dominant component of $e_{\rm st}$).

For $z < 0 \kpc$, the mean vertical velocity driven by the supernovae 
$\langle u_{z} \rangle_{\ell}$ is large and negative, resulting in large, positive
values of $\mu(\rho, \, u_{z}) = \langle\rho u_z\rangle_\ell-
\langle\rho\rangle_\ell\langle u_z\rangle_\ell$.
Thus, the opposite signs of $\langle u_{z} \rangle_{\ell}$ and
$\mu(\rho, \, u_{z})$ result in large, negative values of $e_{\rm st}$ for
negative $z$.
These large, negative values for $e_{\rm st}$ appear to dominate the kinetic energy 
statistics at earlier times. This is discussed in more detail below.

The variation with $\ell$ of the fluctuating kinetic energy produces a more
complicated pattern than for fluctuating magnetic energy, see figure~\ref{fig:e_u}.
The values of $\ell$ for which the variation is weak are $\ell > 300 \p$. 
Such a smoothing length is much larger than any estimate of the correlation 
scale of the random motions, and the optimal smoothing lengths of both $\rho$ or 
$\vect{u}$. As a result, the criterion that the variation of the fluctuating
kinetic energy must be weak is not an appropriate method for choosing suitable 
smoothing lengths for either $\rho$ or $\vect{u}$.

%-------------------------------------------------------------------------------

\section{Influence of the mean-field dynamo}
\label{sect:dynamo}

%-------------------------------------------------------------------------------
\begin{figure*}[tb]
\centering
\includegraphics[width=0.46\textwidth]{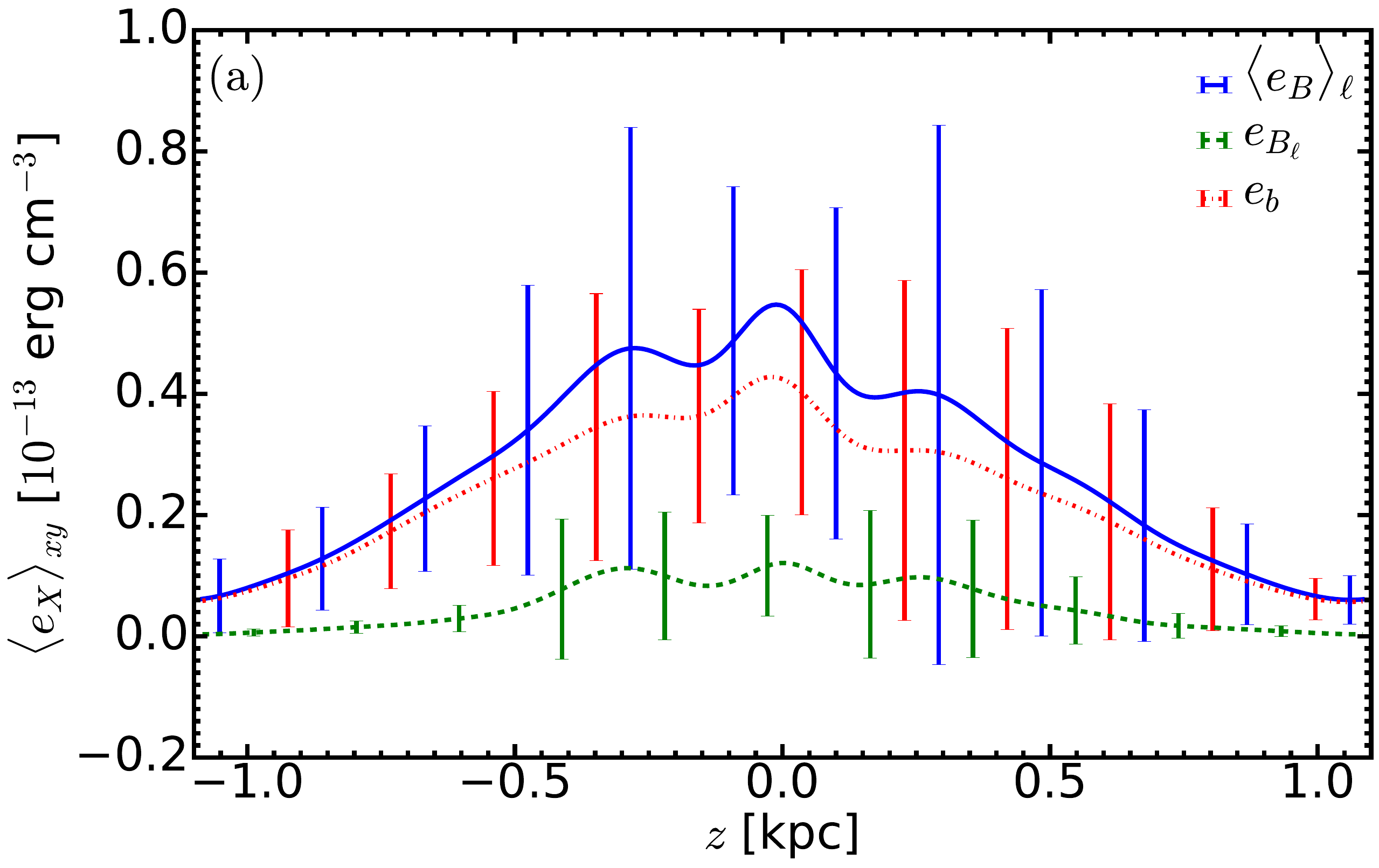}
\\
\includegraphics[width=0.46\textwidth]{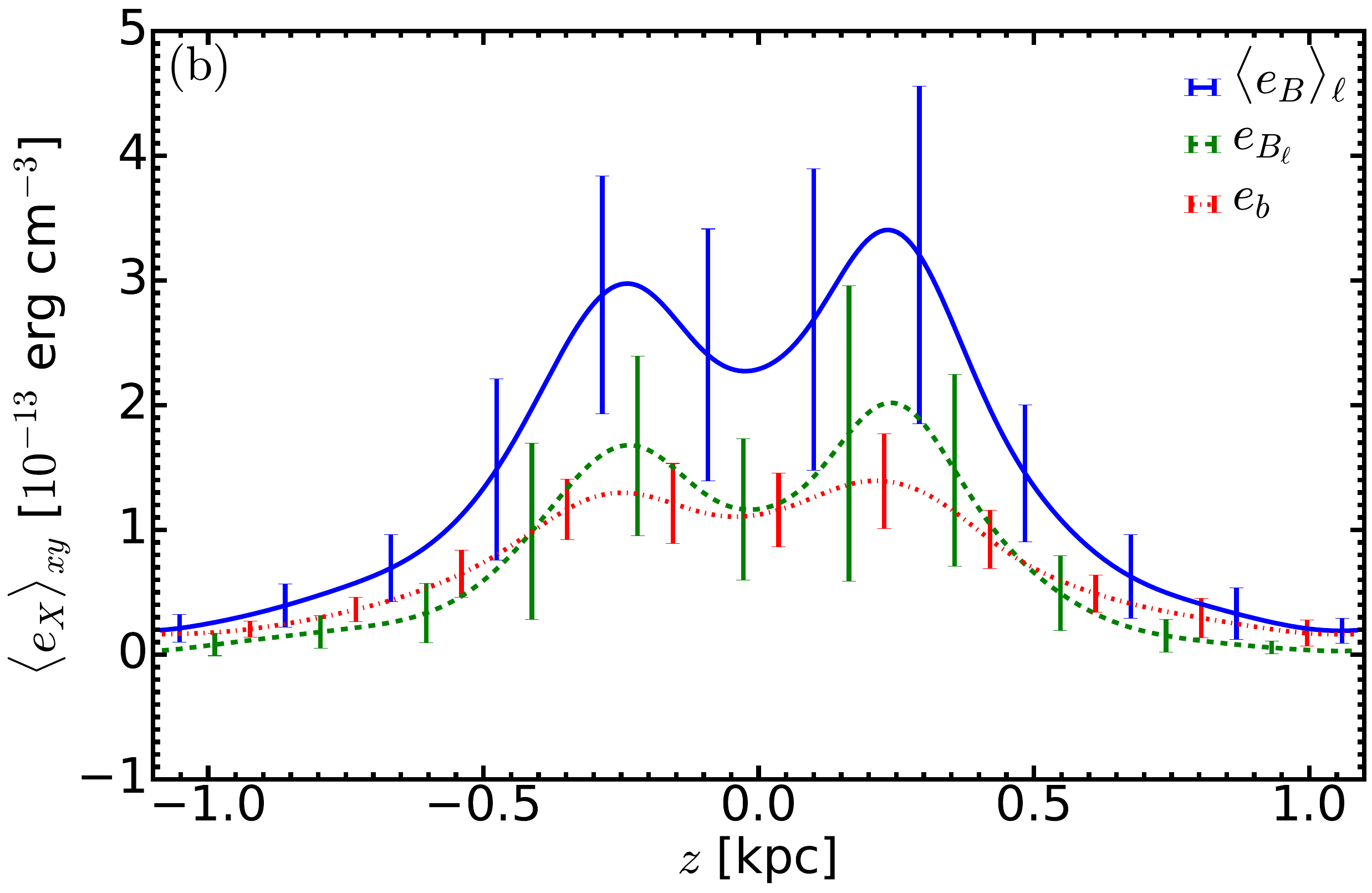}
\\
\includegraphics[width=0.46\textwidth]{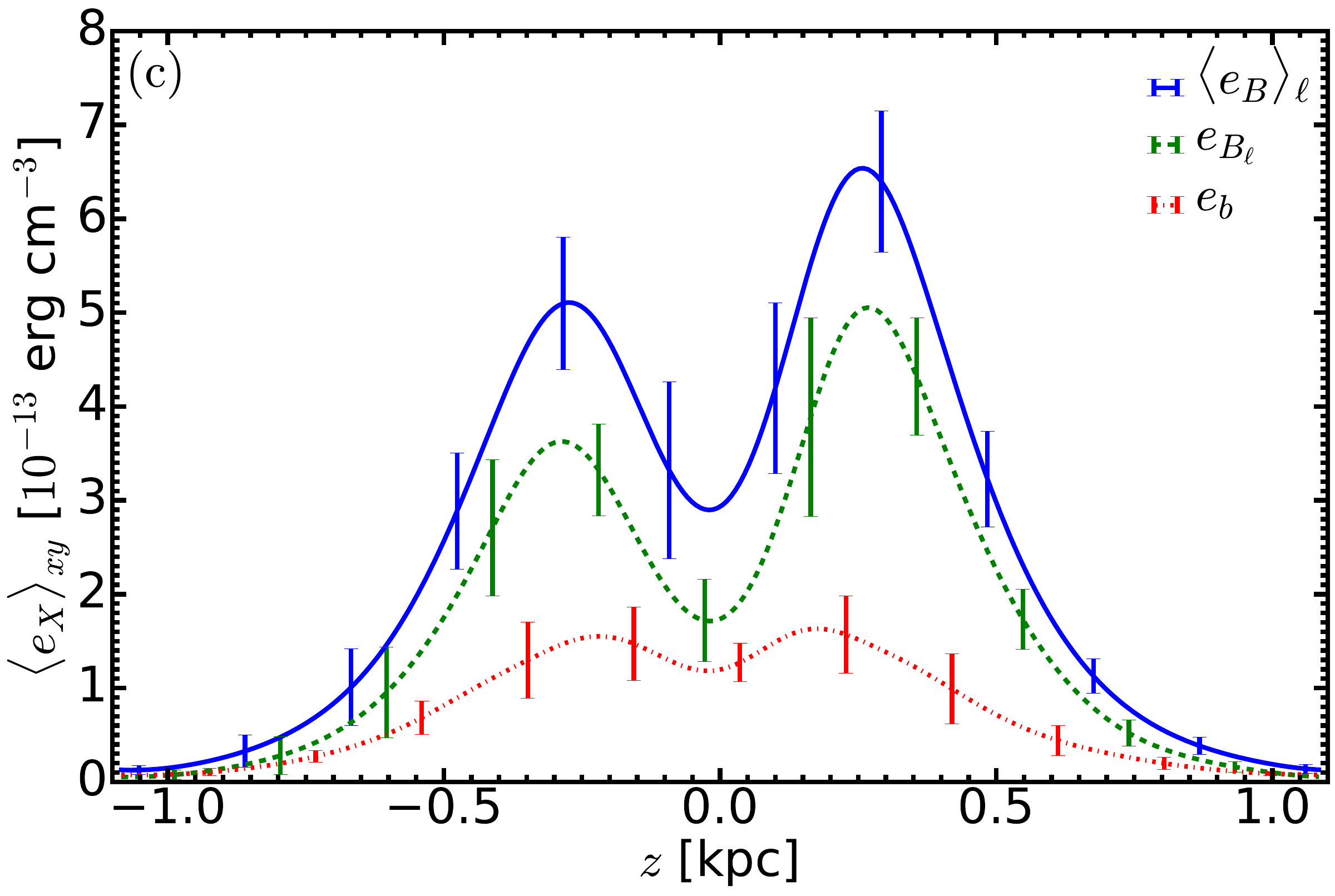}
\caption{Vertical profiles of the horizontal averages of the smoothed total 
magnetic energy, $\langle \langle e_{B} \rangle_{\ell} \rangle_{xy}$ (blue, solid), 
mean magnetic energy, $\langle e_{B_{\ell}} \rangle_{xy}$ (green, dashed), and 
fluctuating magnetic energy, $\langle e_{b} \rangle_{xy}$ (red, dash-dotted); at 
times (a) $0.8 \leq t < 1.1\Gyr$, (b) $1.1 \leq t < 1.45\Gyr$ and 
(c) $t \geq 1.45\Gyr$. The smoothing length applied for each snapshot is 
$\ell = 75 \p$.
}
\label{fig:mag_eng_vertical_profiles}
\end{figure*}
%-------------------------------------------------------------------------------

Figures~\ref{fig:e_b} and \ref{fig:e_u} both suggest that the structures
of magnetic and kinetic energy distributions
vary with the state of the mean-field dynamo. We first examine
the vertical structure of both energies, comparing the three
temporal stages
discussed previously, to demonstrate the changes in structure caused by the dynamo.

\begin{figure*}[tb]
\centering
\includegraphics[width=0.46\textwidth]{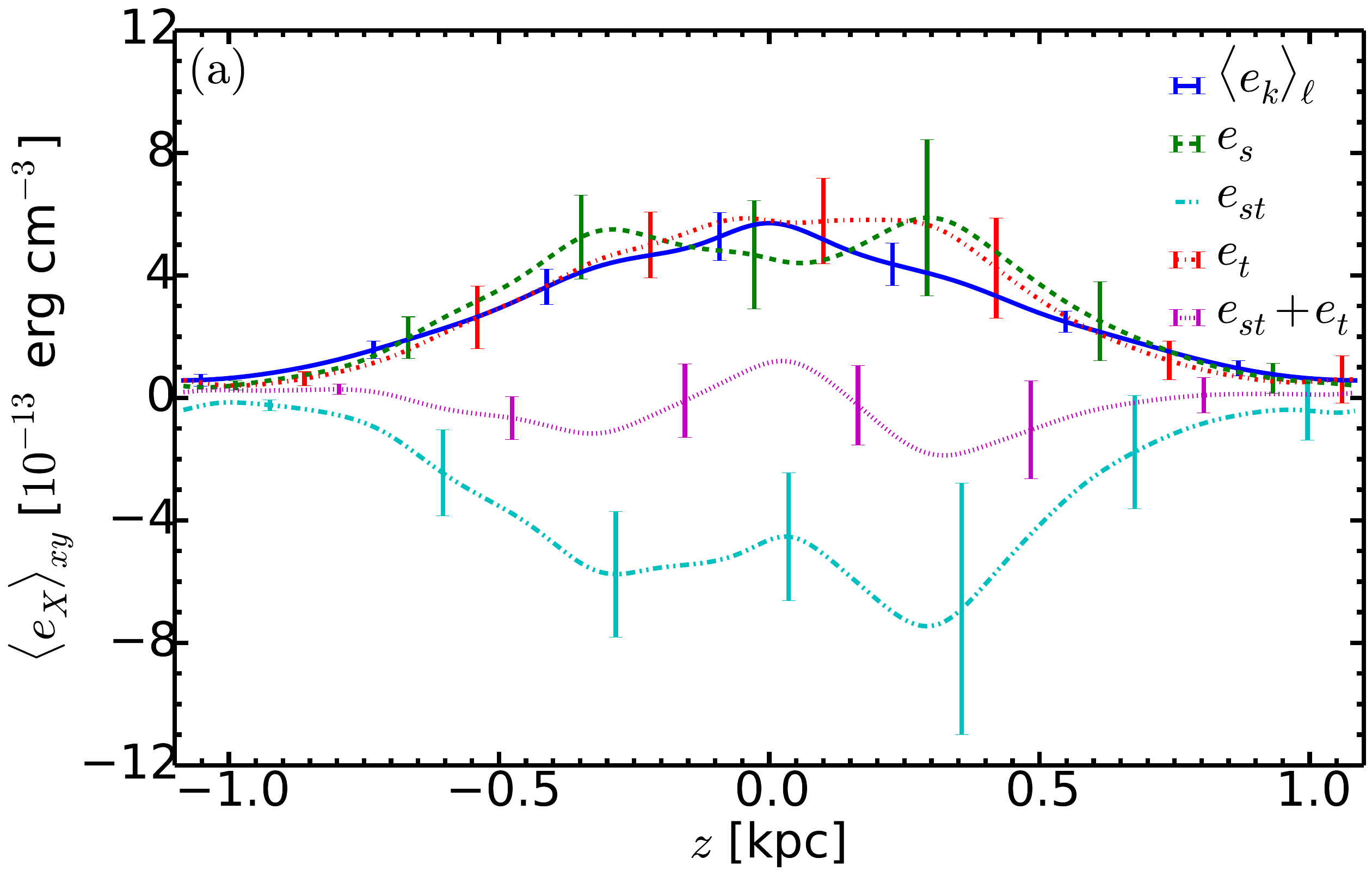}
\\
\includegraphics[width=0.46\textwidth]{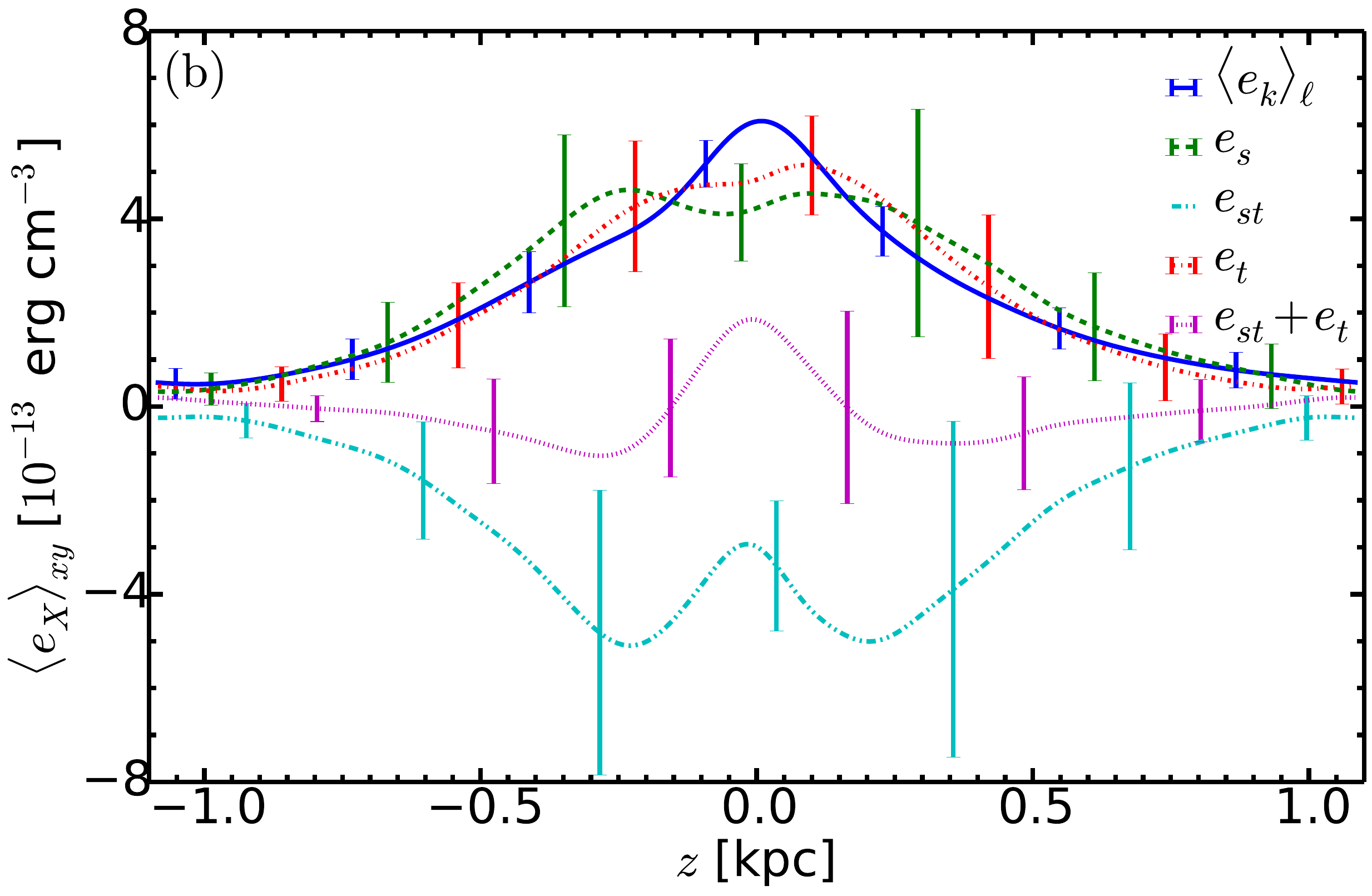}
\\
\includegraphics[width=0.46\textwidth]{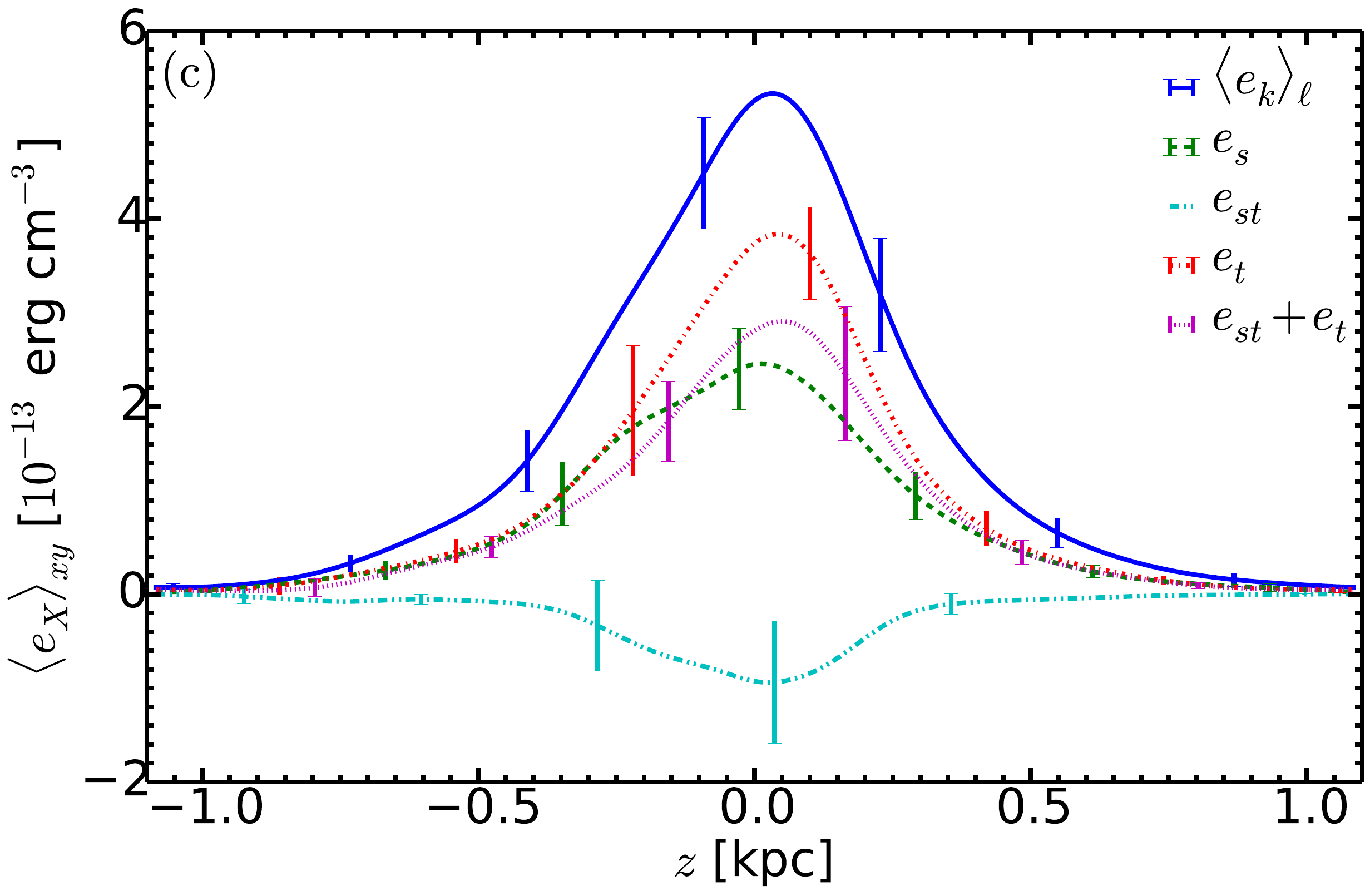}
\caption{As in figure~\ref{fig:mag_eng_vertical_profiles} but for the smoothed total 
kinetic energy density $\langle \langle e_{\rm k} \rangle_{\ell} \rangle_{xy}$
(blue, solid), mean kinetic energy density $\langle e_{\rm s} \rangle_{xy}$ (green, 
dashed), `intermediate scale' kinetic energy density 
$\langle e_{\rm st} \rangle_{xy}$ (cyan, dash-dot-dotted), fluctuating kinetic
energy density $\langle e_{\rm t} \rangle_{xy}$ (red, dot-dashed), and the sum, 
$\langle e_{\rm st} + e_{\rm t} \rangle_{xy}$ (purple, dotted); at times (a) 
$0.8 \leq t < 1.1\Gyr$, (b) $1.1 \leq t < 1.45\Gyr$ and (c) $t \geq 1.45\Gyr$. 
As in figure~\ref{fig:mag_eng_vertical_profiles}, the smoothing length applied 
is $\ell = 75 \p$.
}
\label{fig:kin_eng_vertical_profiles}
\end{figure*}

At early times, when the fluctuating magnetic field dominates the mean field, 
the magnetic field is strongest at $z = 0 \kpc$, 
but significant local maxima (in each of $\langle e_B \rangle_{\ell}$, $e_{B_{\ell}}$ and $e_b$) 
develop at $|z| = 0.3 \kpc$.
This is in line with the kinetic energy, 
where $\langle e_k \rangle_{\ell}$ is maximal at the midplane,
but $e_{\rm s}$ and $e_{\rm st}$ have extrema at $|z| = 0.3 \kpc$;
see figures~\ref{fig:mag_eng_vertical_profiles}a 
and~\ref{fig:kin_eng_vertical_profiles}a. 

%-------------------------------------------------------------------------------
%-------------------------------------------------------------------------------

As the mean field dynamo saturates, the mean magnetic field dominates compared
to the fluctuating field. The vertical profile of the smoothed total magnetic energy 
is increasingly dominated by the mean magnetic energy, 
with the peaks at $|z| = 0.3 \kpc$ now dominating; 
see figures~\ref{fig:mag_eng_vertical_profiles}b,c.
The increasing mean magnetic field significantly alters the vertical profile of
the kinetic energy, as shown in figures~\ref{fig:kin_eng_vertical_profiles}b,c.
All the components in the division of
kinetic energy are ultimately concentrated near the midplane 
(so that the peaks at $z=0.3 \kpc$ have been suppressed),
and the maximum value of $\langle e_{\rm k} \rangle_{\ell}$ decreases. 

Strong mean magnetic fields generated via dynamo action in
the same ISM simulations have been shown to suppress outflows of 
the hot gas \citep[see][]{Evirgen:2017}, which
are associated with high values of kinetic energy. This would lead to a vertical
profile of kinetic energy with the characteristics present in 
figure~\ref{fig:kin_eng_vertical_profiles}c.
The most dramatic change is the effect on the `intermediate scale' component of the
kinetic energy, $e_{\rm st}$. As the magnetic field strength increases, the 
magnitude of the horizontal average of $e_{\rm st}$ decreases significantly, 
becoming small except near to the midplane. 
As a result, the kinetic energy is approximately split between the mean and small-scale energies $e_{\rm s}$ and $e_{\rm t}$. 

As this change appears to be the most significant, we focus on horizontal planes
from the snapshots $t = 0.8 \Gyr$ and $t = 1.6 \Gyr$ at which the vertical 
profiles of $e_{\rm st}$ shown in figure~\ref{fig:kin_eng_vertical_profiles} 
show the most profound differences.

%-------------------------------------------------------------------------------
\begin{figure*}[tb]
\centering
\includegraphics[width=0.92\textwidth]{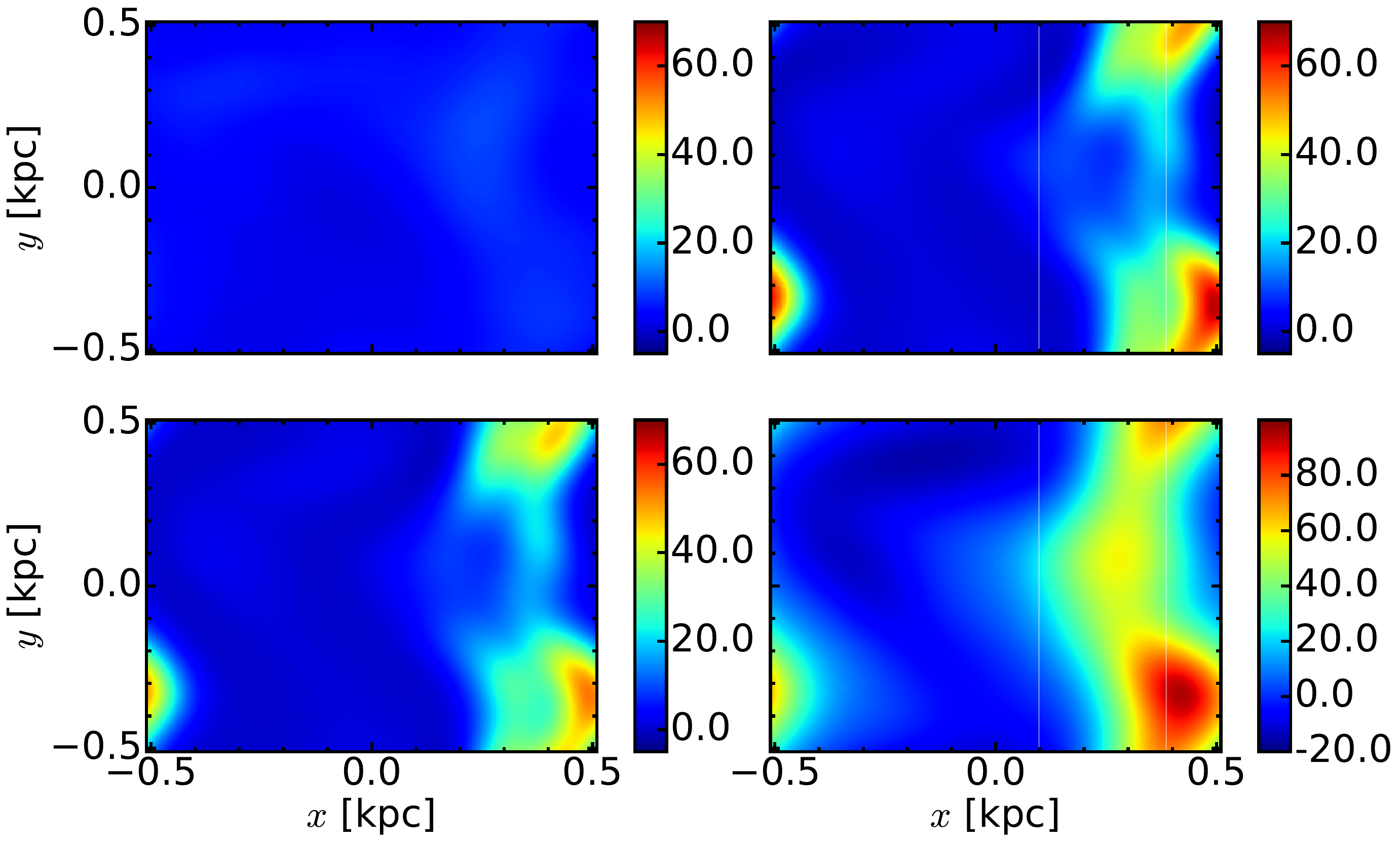}
\caption{Horizontal slices of of the smoothed total kinetic energy density 
$\langle e_{k} \rangle_{\ell}$ (top-left panel), the `intermediate-scale'
kinetic energy density $e_{\rm st}$ (top-right panel), 
$\langle u_{z} \rangle_{\ell} \mu(\rho, \, u_{z})$ the vertical contribution to
$e_{\rm st}$ (bottom-left panel), all in units of $10^{-13} \erg \cm^{-3}$, and
the mean vertical velocity $\langle u_{z} \rangle_{\ell}$ (bottom-right panel)
in $\kms$; at $z=290\p$ from the snapshot $t=0.8\Gyr$. The smoothing length used 
is $\ell=75\p$.}
\label{fig:kin_eng_colourmaps}
\end{figure*}
%-------------------------------------------------------------------------------
 
In the kinematic stage of the mean-field dynamo, there are regions in which 
$e_{\rm st}$ differs from zero significantly,
whilst $\langle e_{k} \rangle_{\ell}$ is 
uniform by comparison (see figure~\ref{fig:kin_eng_colourmaps}).
The mean and turbulent kinetic energies, $e_{\rm s}$ and $e_{\rm t}$ respectively 
(not shown here), are also significant in the same regions as $e_{\rm st}$.
The contribution to $e_{\rm st}$ from the $z$-component of $\vect{u}$,
$\langle u_z\rangle_\ell \, \mu(\rho,u_z)$, comprises a large fraction of the
total kinetic energy (about $80 \%$) and so the vertical flow is dominant 
in $e_{\rm st}$ at this stage.
The values for which $e_{\rm st}$ is highest strongly coincide with regions of large
positive $\langle u_{z} \rangle_{\ell}$, which are the regions of hot gas outflows.
Thus, at the kinematic stage of the mean-field dynamo, $e_{\rm st}$ is strongly 
correlated with the outflows of hot gas.
In this model, the mean magnetic field is absent from the regions of hot gas,
as demonstrated by \citet{Evirgen:2017}.
Thus, the mean magnetic field also avoids regions in which $e_{\rm st}$ 
is significant in magnitude.

\begin{figure*}[tb]
\centering
\includegraphics[width=0.92\textwidth]{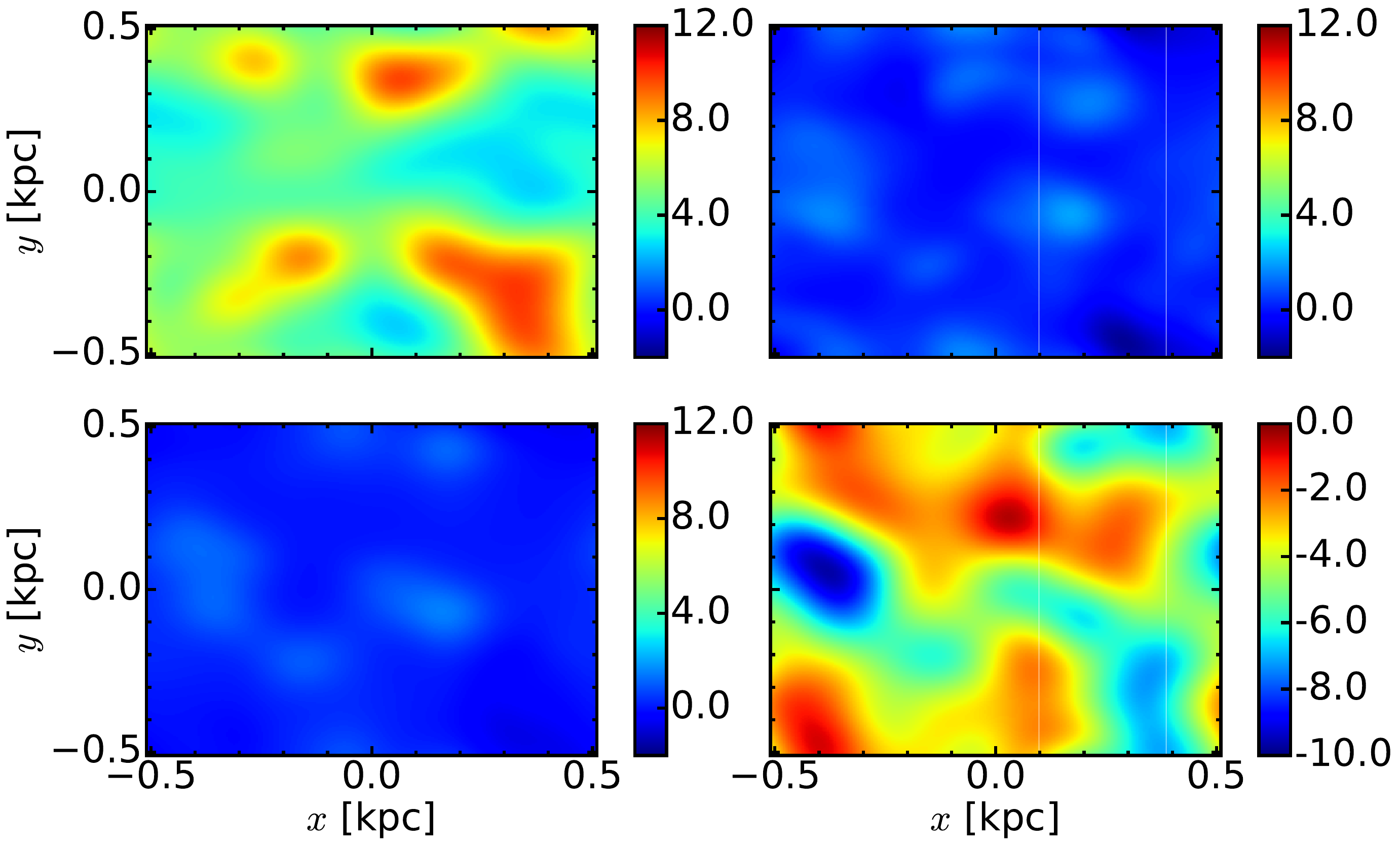}
\caption{As in figure~\ref{fig:kin_eng_colourmaps} but in the plane $z=30\p$, at 
time $t=1.6\Gyr$.}
\label{fig:kin_eng_colourmaps_1}
\end{figure*}

The effect of the amplified mean magnetic field on the kinetic energies is
demonstrated in figure~\ref{fig:kin_eng_colourmaps_1}. The values of $e_{\rm st}$
are reduced significantly and $e_{\rm st}$ appears more uniform.
By contrast, $\langle e_{k} \rangle_{\ell}$ is now more significant and the 
non-uniform structure of $\langle e_{k} \rangle_{\ell}$ is more pronounced.
The contribution to $e_{\rm st}$ from the vertical flow
is also dramatically reduced, and no longer dominates.
The mean vertical velocity is reduced both in maximal value and in the size of
regions in which $\langle u_z\rangle_\ell$ is significant,
indicative of the suppression of the hot gas outflow.
Thus, the partial suppression of the hot gas outflow by the mean magnetic field
has both significantly reduced the value of $e_{\rm st}$ and resulted in the dynamics 
of the overall kinetic energy becoming independent of the behaviour of $e_{\rm st}$.

%-------------------------------------------------------------------------------
%-------------------------------------------------------------------------------

%-------------------------------------------------------------------------------

\section{Comparison with horizontal averaging}
\label{sect:horizontal}

%-------------------------------------------------------------------------------
\begin{figure*}[tb]
  \centering
  \includegraphics[width=0.46\linewidth]{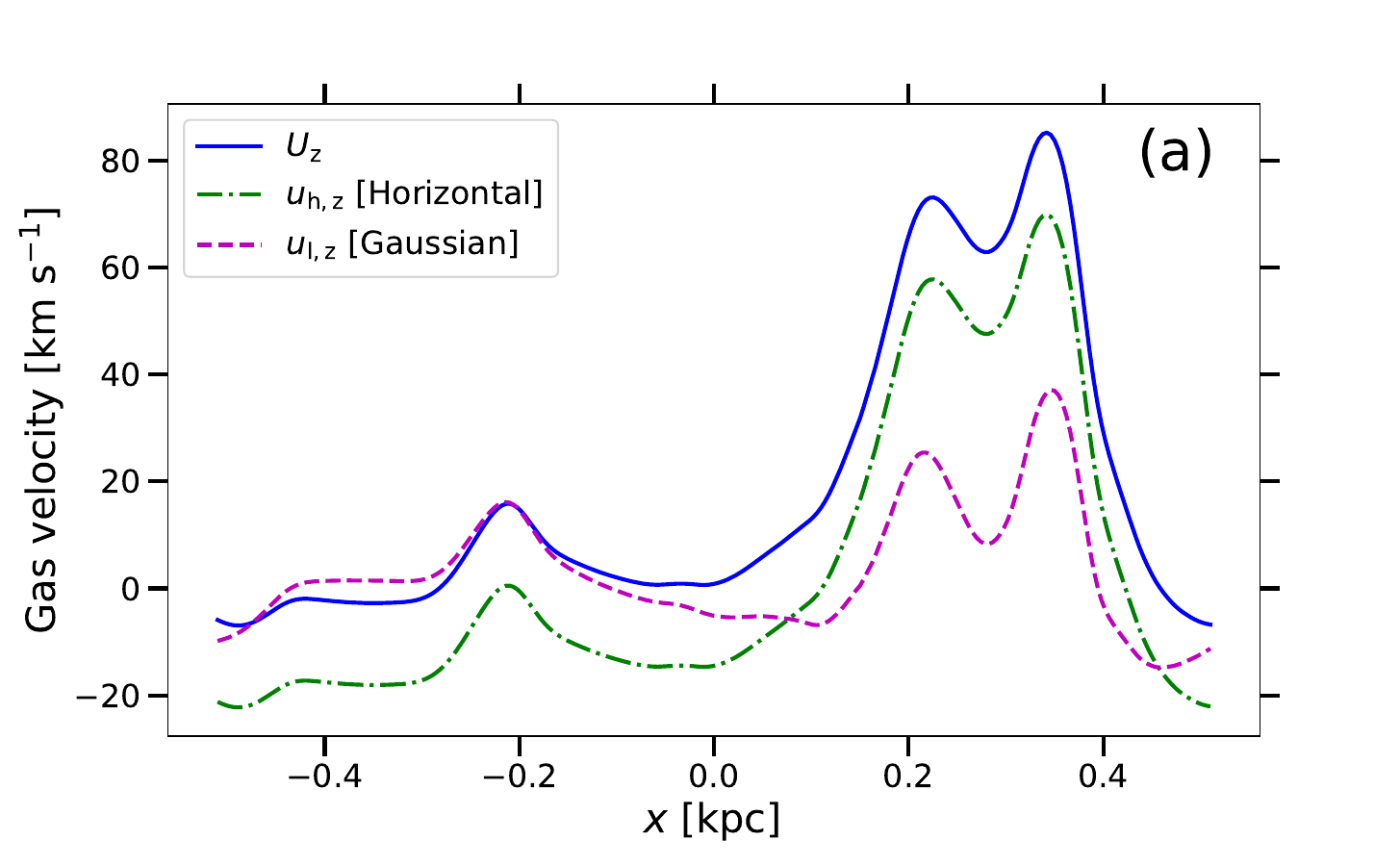}
  \includegraphics[width=0.46\linewidth]{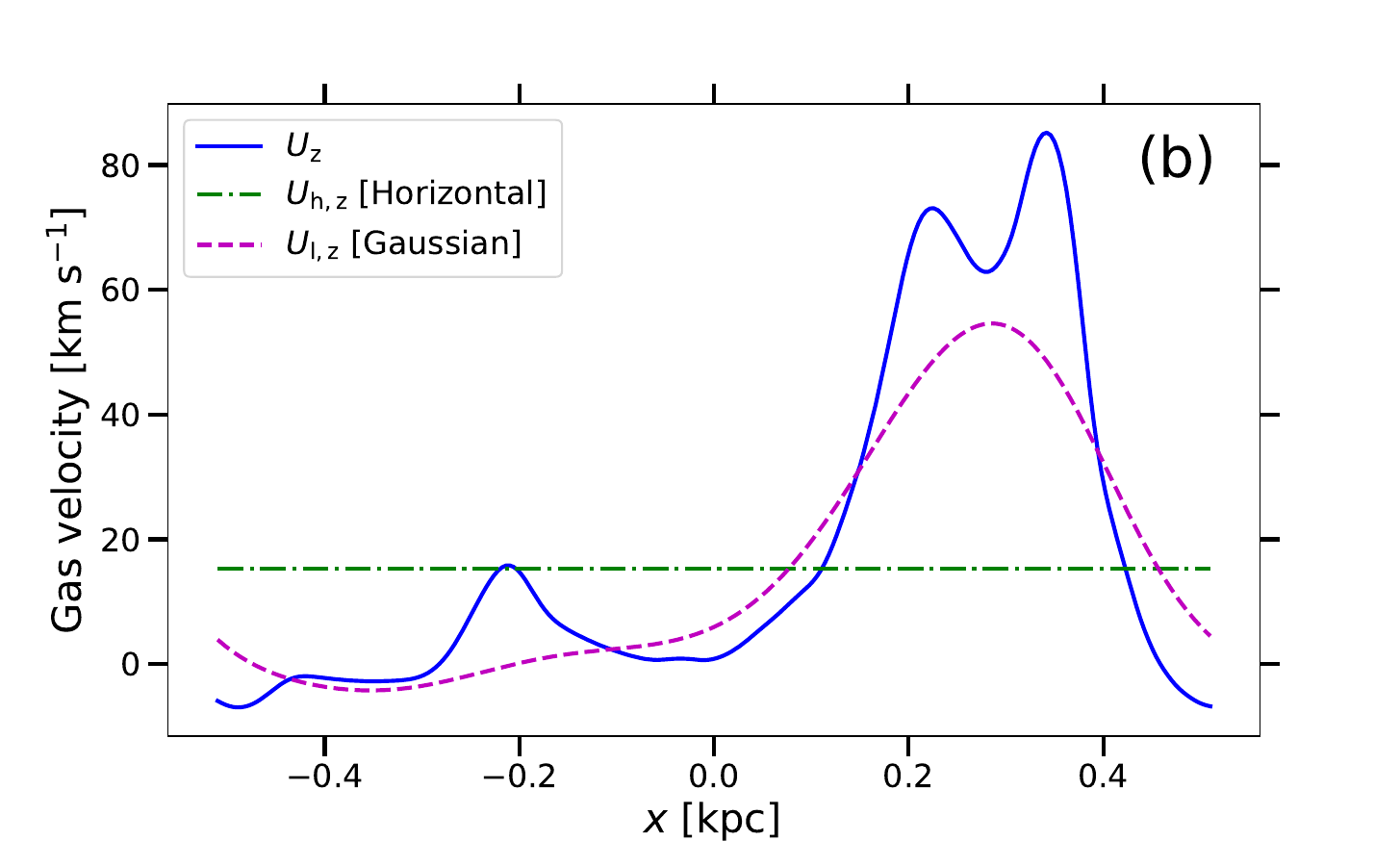}
  \caption{\label{fig:avg_comp}Vertical component of gas velocity as a function of $x$ (galactocentric radius) 
	at fixed $y$ and $z$, 
	to compare horizontal averaging and Gaussian smoothing.
    In Panel~(a), we present the total vertical gas velocity $U_z=u_z$ (blue solid); 
	fluctuating vertical velocity defined using horizontal averaging (green dash dotted), 
	$\rsub{u}{h,z}=u_z-\langle u_z \rangle_{xy}$;
	fluctuating vertical velocity defined using Gaussian smoothing
	(magenta dashed), $\rsub{u}{l,z}=u'_z=u_z-\langle u_z \rangle_{\ell}$.
	In Panel~(b), we present the total vertical gas velocity (blue solid); mean vertical velocity
	defined using horizontal averaging (green dash dotted), 
	$\rsub{U}{h,z}=\langle u_z \rangle_{xy}$;
	mean vertical velocity defined using Gaussian smoothing
	(magenta dashed), $\rsub{U}{l,z}=\langle u_z \rangle_{\ell}$.}
\end{figure*}
%-------------------------------------------------------------------------------

To emphasise the physical relevance of the filtered quantities
as representations of the mean and fluctuating fields,
as compared with the corresponding quantities from horizontal averaging,
in this section we perform a direct comparison.

Figure~\ref{fig:avg_comp} shows a representative comparison of 
%horizontal averaging and Gaussian smoothing.
the two decompositions.
In both panels, 
the solid blue curve gives the total vertical velocity 
at fixed $y$ and $z$;
panel~(a) also shows the fluctuating parts obtained via both averaging methods,
and panel~(b) shows the corresponding mean parts.
We notice that here the velocity is small in magnitude for $x<0$. 
However, there is a region of higher vertical velocity for $x>0$, 
which corresponds to gas flowing away from the midplane. 
This region has the typical characteristics of a hot gas structure. 

Since horizontal averaging takes an average value 
for the entire horizontal plane,
panel~(b) shows the mean vertical gas velocity as constant with respect to $x$,
with $\rsub{U}{h,z}=\langle u_z \rangle_{xy} \approx 20\kms$.
The systematic flow seen in $x>0\kpc$ has approximate velocity 60--80$\kms$.
Consequently, the fluctuating field, derived using horizontal averaging,
retains a large proportion of this systematic flow. 
This is apparent in the similarity of the curves and magnitudes 
of the total vertical velocity and fluctuating vertical velocity 
defined using horizontal averaging.
In general, where there are significant local deviations within a horizontal layer
(as with the localised hot gas structures with large vertical velocity here),
horizontal averaging will not capture the systematic features well.
The horizontal mean calculated may be unrepresentative of the flow in most of the layer
(as in panel~(b)),
and a significant proportion of the fluctuating part may simply be correcting for this locally inappropriate mean (as in panel~(a)), rather than accurately reflecting local fluctuations.
Here this leads to downwards fluctuating flows of up to $20\kms$;
in addition to inflating the magnitude of the fluctuating velocity, 
this misleadingly creates the impression of a systematic local flow towards the midplane.

By contrast, the mean vertical velocity obtained with 
Gaussian smoothing, 
shown by the dashed magenta curve of panel~(b),
extracts the systematic flow well, 
retaining information on the region of hot gas moving away from the plane. 
In addition, 
this mean retains features of the systematic vertical velocity more accurately
in regions where that velocity is smaller (close to zero).
The fluctuating vertical velocity derived using
Gaussian smoothing
retains a smaller proportion of the systematic background trend,
so its magnitude remains commensurately small.

Given that the system discussed here, like many other astrophysical systems,
contains systematic features such as those discussed above ---
driven by the bubbles of hot gas associated with the SN activity in the case of the ISM,
which are well-defined local 3D features that cannot be well represented 
by a horizontal average across the whole layer ---
it is perhaps not surprising that
Gaussian smoothing extracts 
such systematic features more successfully than 
horizontal averaging.
This will also be true in many other instances with pronounced local 3D structures.
(It should be noted that the effect will not be so pronounced where the horizontal
layer is large enough to contain multiple such structures --- involving both
outflows and inflows --- 
which might then average out over the full layer.
But even in such cases, 
the horizontally-averaged mean cannot capture these systematic features,
which should sensibly be characterised as part of the large-scale field, 
rather than as fluctuations.)

Gaussian smoothing (or filtering with any other kernel)
is not technically or conceptually more difficult to implement and interpret 
than any other averaging method, but it requires an appropriate filtering scale to be identified.
The Gaussian kernel
used here is isotropic but the filtering approach also offers the opportunity to use anisotropic kernels wherever appropriate. This can be specially important in the case of the velocity field; for example, SN remnants expand more strongly along the density gradient in galactic discs, and the hot gas is buoyant. It is reasonable to expect that the magnetic field reflects such features of the gas flow and application of an anisotropic filtering kernel might lead to a clearer physical picture, although this would introduce an additional parameter, the degree of the kernel anisotropy. This situation is quite similar to that with the anisotropic wavelet decomposition \citep[e.g.,][]{PFSBBFH06}.

%-------------------------------------------------------------------------------

\section{Discussion}
\label{sect:discussion}

We have applied Gaussian smoothing to obtain mean fields for magnetic field,
density and velocity
in a simulation of the multiphase interstellar gas: a complex, partially ordered magnetohydrodynamic system that supports the mean-field dynamo action.
The optimal smoothing lengths were obtained by spectral 
analysis of each field independently. 
We find $\ell = 75 \p$, approximately $19$ grid cells, 
as an appropriate smoothing length to use for each of these fields.
Such a result is likely to be
sensitive to the choice of simulation parameters,
and should be investigated with a subsequent detailed exploration of the parameter space. 
However, we have successfully demonstrated that an optimal value of $\ell$ 
for each of the three fundamental physical fields is indeed possible 
in our simulation of the ISM.
We have also shown that Gaussian smoothing, 
unlike horizontal averaging,
retains large-scale 3D features of the mean fields.
The filtering approach allows for
a more physically-meaningful decomposition into mean and fluctuating parts
for each variable and for their higher statistical moments, 
such as the magnetic and kinetic energy densities.

It is natural to expect that  the magnetic, density and velocity fields can have distinct spatial structures
since they are controlled by different physical 
processes, even though they do not evolve independently. 
The Gaussian smoothing successfully reveals such differences
(e.g., the different quantities have different correlations lengths),
as shown in \citet{Hollins:2017} for a filtering length of $50 \p$.
(The change to $75 \p$ suggested here does not affect this significantly.)
\citet{GE20} apply kernel filtering
to acquire the mean magnetic field from simulations of SN driven turbulence
and explore
the scale separation between mean and random fields
in their simulations.
They show that the diamagnetic transport of the mean field
is stronger when the filtering scale is smaller.

We examine the mean and fluctuating magnetic and kinetic energies, 
using the generalised central moments
to define the energy density of the fluctuations.
We examine the dependencies of the energies on $\ell$, 
and on the magnetic dynamo saturation.
This allows us to identify the key physical processes 
affecting the mean and fluctuating fields.
Amplification of the mean magnetic field by 
dynamo action has a significant impact on the 
contributions to the magnetic and kinetic energies
due to the mean fields, fluctuations and, in the case of the kinetic energy, 
advection of the fluctuations by the mean flow.
The growing mean magnetic field 
shifts the maximum of the vertical profile of kinetic energy 
towards the disc midplane. 
part of the kinetic energy
density associated with the advection of the velocity fluctuations by the mean flow, $e_{\rm st}$, is closely correlated 
with a systematic gas outflow and is
partly suppressed by the growing mean magnetic field.
This results in a dramatic reduction in $e_{\rm st}$ at late times in the 
simulation, when the kinetic energy 
is mostly associated with the large-scale flow and the velocity fluctuations.

%-------------------------------------------------------------------------------

\section*{Acknowledgement}
The authors wish to acknowledge CSC--IT Center for Science, Finland, for
computational resources Grand Challenge SNDYN.
A.S., A.F., and G.R.S. were supported by the Leverhulme Trust Grant RPG-2014-427 
and STFC Grant ST/N000900/1 (Project 2).
F.A.G.~acknowledges support from the Academy of Finland
ReSoLVE Centre of Excellence (grant 307411) and Ministry
of Education and Culture Global AI-Plasma Physics Pilot.

\bibliographystyle{gGAF}
\bibliography{gaussian_smoothing_refs}
\vspace{12pt}

%-------------------------------------------------------------------------------

\appendices

%-------------------------------------------------------------------------------

\section{Parameters of the numerical model}\label{sect:parameters}
The model discussed here aims to reproduce the statistical properties of 
the random ISM. With the integral scale of random fluctuations in various
physical variables of order $50\p$ \citep{Hollins:2017}, the computational 
domain that we use contains about $400$ correlation cells in each horizontal slice, 
providing sufficient statistics to obtain useful results. Other simulations of 
comparable physical content \citep[e.g.,][]{Hill:2012,Bendre:2015} have 
computational boxes of a similar horizontal size of $0.8\text{--}1\kpc$. 
The next largest physically distinct objects are 
superbubbles, of order $0.5\text{--}1\kpc$ in size,
and OB associations and spiral arms whose scale is of order
$1\text{--}3\kpc$; modelling these phenomena would require significantly larger 
computational domains (and the next generation
of computational models) although some of their features can be captured with
existing models \citep[e.g.,][]{Shukurov:2004,deAvillez:2007}.

The vertical size of the domain is largely controlled by its horizontal 
size. A vertical extent of about $1\kpc$ is insufficient to capture 
fountain flows and model the temperature distribution in the halo, which would 
require heights 
 greater than $5\kpc$ \citep[see][]{Hill:2012}.
However, our simulations are able to fulfil our 
purpose of capturing the physics of the ISM near the midplane, excluding 
fountain flows, without any artefacts from the 
periodic boundary conditions.
As argued by \citet{Gent:2013a}, periodic boundary conditions in the
horizontal planes affect the outflow speed significantly at altitudes exceeding
the horizontal extent of the region. Furthermore, the diameter of supernova 
shells increases to $0.4\text{--}0.6\kpc$ at $|z|\simeq1\kpc$. Therefore, 
results obtained at $|z|\gtrsim1\kpc$ in a computational box of 
$1\times1\kpc^2$ horizontally may be questionable. Results from recent 
simulations performed in computational boxes taller than $1\kpc$ are mostly 
reported only within a few kiloparsec from the midplane 
\citep[e.g.,][]{Hill:2012}. The domain used in our simulations includes two 
scale heights of the warm neutral gas.

With the range of $|z|$ limited to $1\kpc$ 
in our simulations, we have made special effort to ensure that the boundary
conditions at the top and bottom of the domain do not introduce any apparent 
artefacts into numerical solutions, such as a boundary layer with a strong
gradient in any of the physical variables \citep[Appendix~C of][]{Gent:2013a}. 
The limited vertical extent of the box is the main limitation
of our model, but it can only be sensibly increased together with its horizontal size.

The mass loss rate through the top and bottom boundaries is about 
$10^{-3}\Msol\yr^{-1}$,	 so $10^{6}\Msol$ is lost in $1\Gyr$, as compared to 
the total gas mass of $10^{7}\Msol$ in the computational domain. This mass loss 
would correspond to a realistic value of the total mass loss rate of 
$1\Msol\yr^{-1}$ for a galactic disc
of radius $15\kpc$, assuming the Galaxy is 
in a steady state. Our open boundary conditions allow for inflow as well as 
outflow (albeit in a rather \textit{ad hoc} manner),
which mitigates mass loss through the 
boundaries. The mass loss, despite being only modest, was compensated by a 
continuous mass replenishment (in proportion to the local gas density, for 
minimal impact on the dynamics) to maintain an approximately constant gas mass 
throughout the simulations.

The numerical resolution of $4\p$ that we use has been carefully selected to 
reproduce accurately the known expansion laws and approximate internal 
structure of an isolated supernova remnant, subject to radiative cooling 
processes, until its expansion slows down to match the ambient  speed of sound 
\citep[Appendix~B of][]{Gent:2013a}. Thus, we are
confident that our simulations model reliably the associated energy injection 
into the diffuse ISM. Indeed, the intensity of random flows, of order 
$10\kms$ in the warm gas and higher in the hot phase,
is in full agreement with both observations and simulations at a higher
resolution. This is also true of the scales of the random flows, fractional 
volumes of the ISM phases and other aspects of the modelled ISM. 
We have adjusted thermal conductivity so as to ensure that any
structures produced by thermal instability are 
fully resolved at the $4\p$ resolution. 
Comparable simulations of \citet{deAvillez:2007,deAvillez:2012a} have an
adaptive mesh with the finest separation of $1.25\p$, whereas
\citet{Hill:2012} have a resolution of $2\p$, both representing 
an arguably modest improvement.
\citet{GMKS21} have now demonstrated that the small-scale 
(fluctuation) dynamo
solutions are convergent only for a
resolution of
$1\p$ and better, so at $4\p$ the
small-scale dynamo is not sufficiently resolved.
This does not undermine the validity of averaging
procedures presented here, but may
affect the optimal value of the filtering scale $\ell$.
We were unable to identify any further differences in the relevant results of these
simulations that might be a consequence of the difference in
the numerical resolution.

Self-gravity is ignored in our simulations since we do not attempt to model
the very cold molecular gas which is the component significantly affected by 
self-gravity. Simulations with higher resolution would be required to model the 
higher densities and the associated more intense thermal and gravitational 
instabilities.

%-------------------------------------------------------------------------------

\allowdisplaybreaks
\section{Integral forms of the central moments of the second and third order}
\label{IFSMSTO}

The central second-order statistical moment representing the energy density of 
magnetic field fluctuations $e_b$ under smoothing at a scale $\ell$
with a kernel $G_\ell(\vect{x}-\vect{x}')$, with 
$\int_V G_\ell(\vect{x}-\vect{x}')\,\dd\vect{x}'=1$ and 
$X_\ell\equiv\mean{X}_\ell
=\int_V X(\vect{x}') G_\ell(\vect{x}-\vect{x}')\,\dd^3\vect{x}'$ for a scalar 
or vectorial quantity $X$, is given by
\begin{align}
8\pi e_b &=8\pi\left(\mean{e_B}_\ell - e_{B_\ell}\right)
= \mu(b_i, b_i)
= \mean{\vect{B}\cdot\vect{B}}_{\ell} - 
\mean{\vect{B}}_{\ell}\cdot\mean{\vect{B}}_{\ell}     
= \int_V B^{2}(\vect{x}') G_\ell(\vect{x}-\vect{x}')\,\dd^{3}\vect{x}'
- B_\ell^2(\vect{x})\nonumber \\
&=\int_{V} |\vect{B}(\vect{x}') - \vect{B}_\ell(\vect{x})|^2
G_\ell(\vect{x} - \vect{x}') \, \dd^3\vect{x}'
+ 2\int_V \vect{B}(\vect{x}') \cdot \vect{B}_\ell(\vect{x})
G_\ell(\vect{x} - \vect{x}')\,\dd^3\vect{x}' -2B_\ell^2\nonumber \\
&=\int_V |\vect{B}(\vect{x}') -  \vect{B}_\ell(\vect{x})|^2 
G_\ell(\vect{x}-\vect{x}')\,\dd^3\vect{x}'\,.
\label{eq:BiBi}
\end{align}
In a compressible flow, fluctuations in kinetic energy density involve 
second-order statistical moments evaluated as follows:
\begin{align}
\mu(\rho, \vect{u})
&= \mean{\rho\vect{u}}_\ell -\rho_\ell \vect{u}_\ell=
\int_{V} \rho(\vect{x}') 
\vect{u}(\vect{x}') G_\ell(\vect{x}-\vect{x}')\,\dd^{3} \vect{x}' 
      - \rho_\ell(\vect{x}) \vect{u}_{\ell}(\vect{x}) 
\nonumber\\
&=
\int_V \left[\rho(\vect{x}') \, 
      \vect{u}(\vect{x}') - \rho_{\ell}(\vect{x}) 
      \vect{u}_{\ell}(\vect{x})\right] G_{\ell} (\vect{x} - 
      \vect{x}') \, \dd^{3} \vect{x}' 
%\bigg )
\nonumber\\
&=
\int_V \left\{ \left[\rho(\vect{x}') - \rho_\ell(\vect{x})\right]
      \left[\vect{u}(\vect{x}') - \vect{u}_\ell(\vect{x})\right] 
      + \rho(\vect{x}') \vect{u}_\ell(\vect{x})
      + \rho_\ell(\vect{x}) \vect{u}(\vect{x}')  
      - 2\rho_\ell(\vect{x}) \vect{u}_\ell(\vect{x}) \right\} 
      G_\ell(\vect{x} - \vect{x}') \, \dd^3\vect{x}' 
\nonumber \\
&=
\int_V \left[\rho(\vect{x}') - \rho_\ell(\vect{x})\right]
\left[\vect{u}(\vect{x}') - \vect{u}_\ell(\vect{x})\right] 
G_\ell(\vect{x}-\vect{x}') \, \dd^3\vect{x}'
+ \vect{u}_\ell(\vect{x}) \rho_\ell(\vect{x}) + \rho_\ell(\vect{x}) 
\vect{u}_\ell(\vect{x}) - 2 \rho_\ell(\vect{x}) \vect{u}_\ell(\vect{x})
\nonumber \\
&=\int_V \left[\rho(\vect{x}') - \rho_\ell(\vect{x})\right]
\left[\vect{u}(\vect{x}') - \vect{u}_\ell(\vect{x})\right] 
G_\ell(\vect{x}-\vect{x}') \, \dd^3\vect{x}'\,.
\label{eq:rhoui}
\end{align}
Similarly to equation~\eqref{eq:BiBi},
\begin{align}
\mu(u_i, u_i)  &= \int_V |\vect{u}(\vect{x}') - \vect{u}_\ell(\vect{x})|^2
G_{\ell}(\vect{x}-\vect{x}')\,\dd^3\vect{x}'\,.
\label{eq:uiui}
\end{align}
The third-order moment that appears in the kinetic energy density
follows as (summation over repeated indices is understood)
\begin{align}
\mu(\rho, u_i, u_i)
&=\mean{\rho u_i u_i}_\ell - 2\mean{u_i}_\ell\, \mu(\rho,u_i) - 
\mean{\rho}_\ell\, \mu(u_i,u_i) - \mean{\rho}_{\ell} \mean{u_i}_\ell 
\mean{u_i}_\ell\nonumber\\
&=\mean{\rho u_i u_i}_\ell - 2\mean{u_i}_\ell \Bigl(\mean{\rho u_i}_\ell - 
\mean{\rho}_{\ell} \mean{u_i}_\ell\Bigr) 
- \mean{\rho}_\ell\Bigl(
\mean{u_i u_i}_\ell-\mean{u_i}_\ell \mean{u_i}_\ell\Bigr)
- \mean{\rho}_\ell \mean{u_i}_\ell \mean{u_i}_\ell\nonumber \\
&=\mean{\rho u_i u_i}_\ell - 2\mean{u_i}_\ell\mean{\rho u_i}_\ell -  
\mean{\rho}_\ell\mean{u_i u_i}_\ell
+2\mean{\rho}_\ell\mean{u_i}_\ell\mean{u_i}_\ell\nonumber\\
&=\int_V \left[ \rho(\vect{x}') u^2(\vect{x}') 
- 2\vect{u}_\ell(\vect{x}) \cdot \rho(\vect{x}')\vect{u}(\vect{x}') 
- \rho_\ell(\vect{x}) u^2(\vect{x}')
+ 2\rho_\ell(\vect{x}) u_\ell^2(\vect{x}) \right]
G_\ell(\vect{x}-\vect{x}')\,\dd^3\vect{x}' 
\nonumber\\ 
&=\int_V \left[\rho(\vect{x}') - \rho_\ell(\vect{x})\right] 
\left|\vect{u}(\vect{x}') - \vect{u}_\ell(\vect{x})\right|^2
G_\ell(\vect{x}-\vect{x}')\,\dd^3\vect{x}' - u_\ell^2(\vect{x}) 
\int_V \rho(\vect{x}') G_\ell(\vect{x}-\vect{x}') \,\dd^3\vect{x}' 
\nonumber\\
&\mbox{}\qquad
- 2\rho_\ell(\vect{x}) \vect{u}_\ell(\vect{x}) 
\cdot \int_V \vect{u}(\vect{x}')G_\ell(\vect{x}-\vect{x}')\,\dd^3\vect{x}'      
+3\rho_\ell(\vect{x})u_\ell^2(\vect{x}) 
\nonumber\\
&=\int_V \left[\rho(\vect{x}') - \rho_\ell(\vect{x})\right]
\left|\vect{u}(\vect{x}') - \vect{u}_\ell(\vect{x})\right|^2
G_\ell(\vect{x}-\vect{x}')\,\dd^3\vect{x}'\,.
\label{eq:rhouiui}
\end{align}

%-------------------------------------------------------------------------------

\end{document}